\documentclass[preprint,10pt,authoryear]{elsarticle}
\usepackage[margin=0.5in]{geometry}

\usepackage{hyperref}
\usepackage{astrojournals}
\usepackage{amsmath}
\usepackage{graphicx}
\usepackage{txfonts}
\usepackage{color}
\usepackage[percent]{overpic}

\journal{New Astronomy}

\begin{document}

\begin{frontmatter}

  \title{Accretion discs with non-zero central torque}

  \author[label1]{C.~J.~Nixon\corref{cor1}} \ead{cjn@leicester.ac.uk}
  \author[label1,label2]{J.~E.~Pringle}
  \cortext[cor1]{Corresponding author}
  \address[label1]{School of Physics and Astronomy, University of Leicester, Leicester, LE1 7RH, UK}
  \address[label2]{Institute of Astronomy, Madingley Road, Cambridge, CB3 0HA, UK}

\begin{abstract}
  We present analytical and numerical solutions for accretion discs subject to a non-zero central torque. We express this in terms of a single parameter, $f$, which is the ratio of outward viscous flux of angular momentum from the inner boundary to the inward advected flux of angular momentum there. The standard ``accretion'' disc, where the central boundary condition is zero-torque, is represented by $f=0$. A ``decretion'' disc, where the radial velocity at the inner boundary is zero, is represented by $f\rightarrow\infty$. For $f > 0$ a torque is applied to the disc at the inner boundary, which feeds both angular momentum and energy into the disc. This can arise, for example, in the case of a circumbinary disc where resonances transfer energy and angular momentum from the binary to the disc orbits, or where the disc is around a rotating magnetic star which can allow the disc orbits to be accelerated outwards at the magnetospheric radius. We present steady-state solutions to the disc structure as a function of $f$, and for arbitrary kinematic viscosity $\nu$. For time-dependent discs, we solve the equations using a Green's function approach for the specific case of $\nu \propto R$ and provide an example numerical solution to the equations for the case of $\nu \propto R^{3/2}$. We find that for values of $f\lesssim 0.1$ the disc solutions closely resemble ``accretion'' discs. For values of $f \gtrsim 10$ the solutions initially resemble ``decretion'' discs, but at sufficiently late times exhibit the properties of ``accretion'' discs. We discuss the application of this theory to different astrophysical systems, and in particular the values of the $f$ parameter that are expected in different cases.
\end{abstract}


\end{frontmatter}

\section{Introduction}
\label{intro}
In a standard accretion disc, all of the material arriving at the inner radius, $R_{\rm in}$, is accreted by the central object, along with all  its angular momentum. This is equivalent to setting a zero viscous torque boundary condition at $R = R_{\rm in}$. 

For a steadily accreting disc of this kind, as shown by \cite{Shakura:1973aa}, the energy dissipated per unit area is given by
\begin{equation}
\label{dissipation0}
D(R) = \frac{3GM{\dot M}}{4\pi R^3}\left[1-\left(\frac{R_{\rm in}}{R}\right)^{1/2}\right]\,,
\end{equation}
where $M$ is the mass of the central object and ${\dot M}$ is the steady accretion rate through the disc. Thus the total luminosity released in such a disc is
\begin{equation}
\label{E1}
L = \int_{R_{\rm in}}^\infty 2\pi R D(R)\,{\rm d}R = \frac{GM{\dot M}}{2R_{\rm in}}\,.
\end{equation}
This is one half of the available gravitational energy. The other half is advected through the inner boundary in the form of kinetic energy of circular motion. 

The lack of a central torque also implies that all the angular momentum arriving at the inner boundary is also advected inwards. This is a good approximation, for example, for a thin disc around a star which is not rotating close to break up \citep[e.g.][]{Pringle:1981aa}. However, this may not always be the case. Examples of where the inner torque on the disc might not be  zero to a good approximation include:
\begin{enumerate}
\item an accretion disc which is truncated by a stellar magnetosphere,
\item a ``decretion'' disc around a Be star, and
\item a circumbinary disc.
\end{enumerate}
The torque at the inner boundary has also been discussed in the context of the appropriate boundary condition to apply at the innermost stable circular orbit (ISCO) of a disc around a black hole.

Here we explore disc solutions in the presence of a non-zero torque at the inner boundary of the disc. In Section~\ref{eqns} we provide the underlying equations. In Section~\ref{steady} we provide steady state solutions to the disc structure. In Section~\ref{timedep} we provide analytical and numerical solutions to the time-dependent disc evolution. In Section~\ref{systems} we discuss different astrophysical systems for which these models may be appropriate, and we conclude in Section~\ref{conclusions}.

\section{Disc equations}
\label{eqns}
The disc equations describing the time dependence of surface density, $\Sigma(R,t)$ where $R$ is radius and $t$ time,  are  \citep[see, for example,][]{Pringle:1981aa}, first, conservation of mass:
\begin{equation}
\label{mass}
\frac{\partial \Sigma}{\partial t} + \frac{1}{R} \frac{\partial}{\partial R} \left(R\Sigma \varv_R\right) = 0\,,
\end{equation}
where $\varv_R$ is the radial velocity, and, second, conservation of angular momentum
\begin{equation}
\label{angmom}
\frac{\partial}{\partial t}\left(\Sigma R^2 \Omega\right) + \frac{1}{R}\frac{\partial}{\partial R}\left(R^3\Sigma\Omega \varv_R\right) = \frac{1}{R}\frac{\partial}{\partial R}\left(\nu\Sigma R^3\Omega^\prime\right)\,.
\end{equation}
Here $\Omega(R)$ is the angular velocity of the disc material, $\nu$ the vertically averaged kinematic viscosity, and $\Omega^\prime = {\rm d}\Omega/{\rm d}R$.

By manipulating these we find that
\begin{equation}
\label{vR}
\varv_R = \frac{ \frac{\partial}{\partial R} (\nu \Sigma R^3 \Omega^\prime)}{\Sigma R (R^2 \Omega)^\prime}\,.
\end{equation}

Henceforth we specialise to Keplerian discs around a central object of mass $M$, and take $\Omega = \sqrt{GM/R^3}$.

By integrating Equation~\ref{angmom} over the disc, we find that the flux of angular momentum through radius $R$ is given by
\begin{equation}
\label{Hflux}
J = J_{\rm adv} + J_{\rm visc} = 2\pi R\Sigma (GMR)^{1/2}\varv_R + 3\pi\nu\Sigma (GMR)^{1/2}\,.
\end{equation}
Here the first term is the advected flux  (the direction of which depends on the sign of $\varv_R$) and the second term is the viscous flux, which is always directed outwards\footnote{This is true in the body of the disc, where $\Omega \propto R^{-3/2}$, but not, for example, in a boundary layer where we might have ${\rm d}\Omega/{\rm d}R < 0$.}.

Combining Equations~\ref{mass} and~\ref{vR} we obtain the usual equation for the evolution of the surface density (here for a Keplerian disc)
\begin{equation}
\label{Sigma}
\frac{\partial \Sigma}{\partial t} = \frac{3}{R} 
\frac{\partial}{\partial R} \left[ R^{1/2} \frac{\partial}{\partial R} \left( \nu \Sigma R^{1/2} \right)\right] + \frac{\dot M}{2 \pi R_{\rm add}} \delta(R - R_{\rm add}),
\end{equation}
where we have added the possibility of adding mass at a rate $\dot M$ at radius $R_{\rm add}$.

\subsection{Boundary conditions}
We are mainly interested in the inner boundary. There it is usual to take one of two possibilities.

\begin{enumerate}

\item $\Sigma = 0$. This implies (see Equation~\ref{Hflux}) that the viscous torque there is zero. This is appropriate for an accretion disc around a slowly rotating star. All the matter that reaches the inner boundary is accreted along with its angular momentum.

\item $\varv_R = 0$, or equivalently, $\partial (\nu \Sigma R^3 \Omega^\prime) /\partial R = 0$. This implies that there is no accretion at the inner boundary. This is appropriate for a decretion disc. Nothing is accreted at the inner boundary. There is a source of energy and angular momentum at the inner boundary which enables all the disc matter to be expelled to larger radii.

\end{enumerate}

In this paper we are interested in the more general possibility, namely that not all of the angular momentum that  reaches the inner boundary is accreted by the central object. In a Keplerian disc, the advected angular momentum flux is
\begin{equation}
J_{\rm adv} = -6\pi R^{1/2} \left(GMR\right)^{1/2} \frac{\partial}{\partial R} \left(\nu \Sigma R^{1/2}\right)\,,
\end{equation}
and the viscous flux is
\begin{equation}
J_{\rm visc} =  3 \pi \nu \Sigma (GMR)^{1/2}.
\end{equation}

For the general inner boundary condition we shall assume that the outward (viscous) flux of angular momentum at the inner boundary is $f$ times the inward (advected) flux. This is written succinctly in Equation~\ref{innerBC} below. We note for the time being that $f=0$ is the standard accretion disc condition, while for a decretion disc we need to take the limit $f \rightarrow \infty$.

It is also worth noting at this point that this choice of boundary condition ensures that it is a {\em local} condition, applicable at the inner boundary, and dependent solely on local conditions there. Throughout this paper we shall carry out the analysis assuming that $f$ remains a constant, though in practice, depending on the astrophysical application $f$ may vary with time. We discuss such possible scenarios in Section~\ref{systems}.~\footnote{In contrast, \cite{Rafikov:2013aa} in his study of circumbinary discs, chooses to adopt a non-local boundary condition in which the accretion rate at the inner boundary is deemed to be some fraction (less than unity) of the accretion rate at large radius in the disc. Such a choice of boundary condition, in which the inner disc boundary has apparent knowledge of conditions at large radii, poses questions of causality. \cite{Rafikov:2016aa} presents similarity solutions for centrally torqued discs valid for $0 \le R < \infty$, and $t > 0$. For similarity solutions, he must treat the inner boundary as being at $R_{\rm in} = 0$, and thus the rate of advection of angular momentum at the inner boundary is always zero, and the central torque is a prescribed function of time. Thus, for these solutions, the concept of $f$ introduced here is not meaningful.}

\section{Steady disc}
\label{steady}
To illustrate the implications of an effective inner torque, we first consider a steady disc between radii $R_{\rm in}$ and $R_{\rm out}$ with matter being added at a constant rate ${\dot M}$ at a radius $R_{\rm add}$, where $R_{\rm in} < R_{\rm add} < R_{\rm out}$.  We are also interested in the infinite disc, which is the limiting case $R_{\rm out} \rightarrow \infty$. For simplicity we consider the case of a Keplerian disc, for which $\Omega^2 = GM/R^3$. To simplify the algebra we define 
\begin{equation}
S = \nu \Sigma R^{1/2}.
\end{equation}

The usual boundary conditions are either zero torque ($S = 0$) or zero radial velocity (${\rm d}S/{\rm d}R = 0$). However, here we are interested in the case where the inner boundary condition allows an amount of matter to be accreted but not necessarily with all of its angular momentum. This means that there is an outward viscous flux of angular momentum which is a fraction, $f$, of the inward advected flux. In this case, the inner boundary condition is
\begin{equation}
\label{innerBC}
\left[S - 2f R\frac{{\rm d}S}{{\rm d}R}\right]_{R_{\rm in}} = 0\,.
\end{equation}
We note that the standard accretion disc corresponds to $f = 0$, and the decretion disc to the limit $f \rightarrow \infty$, and again we stress that with $f \neq 0$ this is still a local condition~\footnote{We note that this is an example of a Robin boundary condition, as opposed to the Dirichlet and Neumann boundary conditions for the accretion and decretion disc cases respectively \citep[see, for example,][]{Eriksson:1996aa}.}.

At the outer boundary, we shall assume for simplicity that $S=0$, so that all the mass and angular momentum reaching the outer boundary is absorbed there\footnote{This also ensures that there is a steady disc solution, even in the limit $f \rightarrow \infty$. We also note that other choices are possible -- see, for example, \citep{Martin:2011aa}.}

Then the solution to the steady disc equations (eqn~\ref{Sigma} with $\partial\Sigma/\partial t$ = 0) for $R \neq R_{\rm add}$ is
\begin{equation}
S = C R^{1/2} + D,
\end{equation} where $C$ and $D$ are constants.

To allow for addition of mass at $R = R_{\rm add}$ there are jump conditions evaluated at $R = R_{\rm add}$:
\begin{enumerate}
\item $S(R)$ is continuous, so that $S(R_{\rm add}^+) = S(R_{\rm add}^-)$, and 
\item The jump in advected angular momentum flux at $R_{\rm add}$ equals the angular momentum flux from the added material, which can be written as
\begin{equation}
\left[\frac{{\rm d}S}{{\rm d}R}\right]^{R_{\rm add}^+}_{R_{\rm add}^-} =  -\frac{\dot M}{6\pi R_{\rm add}^{1/2}}\,.
\end{equation}
\end{enumerate}

Once we have found $S = \nu \Sigma R^{1/2}$, we can find the energy dissipation rate per unit area $D(R) = \nu\Sigma\left(R\Omega^\prime\right)^2$. Applying the jump conditions and the inner and outer boundary conditions leads to the following solutions:\\
(i) For $R_{\rm in} \le R < R_{\rm add}$, the solution is 
\begin{equation}
\label{s1}
S(R_{\rm in} \le R < R_{\rm add}) = \frac{\dot M}{3\pi}\left[R^{1/2} - (1-f) R_{\rm in}^{1/2}\right]\frac{R_{\rm out}^{1/2} - R_{\rm add}^{1/2}}{R_{\rm out}^{1/2}-(1-f)R_{\rm in}^{1/2}}\,.
\end{equation}
In the limit $R_{\rm out} \rightarrow \infty$ this becomes
\begin{equation}
S = \frac{\dot M}{3\pi}\left[R^{1/2} - (1-f)R_{\rm in}^{1/2}\right]\,.
\end{equation}
and we find
\begin{equation}
\label{modifieddissipation}
D(R) = \frac{3GM{\dot M}}{4\pi R^3}\left[1-(1-f)\left(\frac{R_{\rm in}}{R}\right)^{1/2} \right]\,.
\end{equation}

This is the standard text book result \citep[see, for example,][]{Shapiro:1983aa}. It differs from the usual (zero torque) expression (Equation~\ref{dissipation0}) because extra energy is given to the disc via the inner torque.

(ii) For $R_{\rm add} < R \le R_{\rm out}$, the solution is
\begin{equation}
\label{s2}
S(R_{\rm add} < R \le R_{\rm out}) = \frac{\dot M}{3\pi}\left[R_{\rm add}^{1/2}-(1-f)R_{\rm in}^{1/2}\right]\frac{R_{\rm out}^{1/2}-R^{1/2}}{R_{\rm out}^{1/2}-(1-f) R_{\rm in}^{1/2}}\,.
\end{equation}
Note that at $R=R_{\rm add}$ the solution must be continuous, and by comparing eqns~\ref{s1} \& \ref{s2} at $R=R_{\rm add}$ we can see that this is the case. In the limit $R_{\rm out} \rightarrow \infty$ we find
\begin{equation}
S = \frac{\dot M}{3\pi}\left[R_{\rm add}^{1/2}-(1-f)R_{\rm in}^{1/2}\right]\,,
\end{equation}
and
\begin{equation}
D(R) = \frac{3GM{\dot M}}{4\pi R^3}\left[\frac{R_{\rm add}^{1/2}-(1-f)R_{\rm in}^{1/2}}{R^{1/2}}\right]\,.
\end{equation}

Thus the inner disc acts like a slightly modified accretion disc, and the outer disc like a steady decretion disc (Pringle 1991). 

For a steady accretion disc, coming in from infinite radius (that is, in the limit $R_{\rm add}, R_{\rm out} \rightarrow \infty$), the total energy emitted by the disc is now
\begin{equation}
\label{energy}
L = \int_{R_{\rm in}}^\infty 2\pi R D(R)\,{\rm d}R = \frac{GM{\dot M}}{2R_{\rm in}}(1+2f)\,.
\end{equation}
By comparison with Equation~\ref{E1}, we see that the added torque at the centre increases the total energy emitted by the disc.

For a steady accretion disc, it is evident that if $f > 1/2$, more energy is emitted by the disc than is available from the accreted material. Thus once $f > 1/2$, additional energy must be provided at the inner boundary in order to power the disc. 

Finally, for a steady disc, with a mass input rate $\dot M$ at $R = R_{\rm add}$ we may ask where that mass ends up.  For the inner region $R_{\rm in} < R < R_{\rm add}$ the mass flux is inwards at a rate
\begin{equation}
  \label{mdotin_steady}
{\dot M}_{\rm in} = {\dot M}\left\{\frac{R_{\rm out}^{1/2}-R_{\rm add}^{1/2}}{R_{\rm out}^{1/2}-(1-f)R_{\rm in}^{1/2}}\right\}\,,
\end{equation}
and for the outer region $R_{\rm add} < R < R_{\rm out}$ the mass flux is outwards at a rate
\begin{equation}
  \label{mdotout_steady}
{\dot M}_{\rm out} = {\dot M}\left\{\frac{R_{\rm add}^{1/2}-(1-f)R_{\rm in}^{1/2}}{R_{\rm out}^{1/2}-(1-f)R_{\rm in}^{1/2}}\right\}\,.
\end{equation}
These expressions imply that we need to take care when we are considering decretion ($f \rightarrow \infty$) discs of infinite extent ($R_{\rm out} \rightarrow \infty$). 
\begin{enumerate}
\item For a disc of infinite extent, with any non-zero rate of mass accretion on to the central object ($f$ finite, {\em however large}, and $R_{\rm out} \rightarrow \infty$) we see that all the mass ends up in the centre, that is ${\dot M}_{\rm in} = {\dot M}$ and ${\dot M}_{\rm out} = 0$,
\item whereas for a disc of fixed radial extent ($R_{\rm out}$ fixed), and a central torque that can prevent any accretion ($f \rightarrow \infty$) we see that all the mass is expelled through the outer boundary, that is ${\dot M}_{\rm in} = 0$ and ${\dot M}_{\rm out} = {\dot M}$.
\end{enumerate}

\begin{figure}
  \centering\includegraphics[width=0.5\columnwidth]{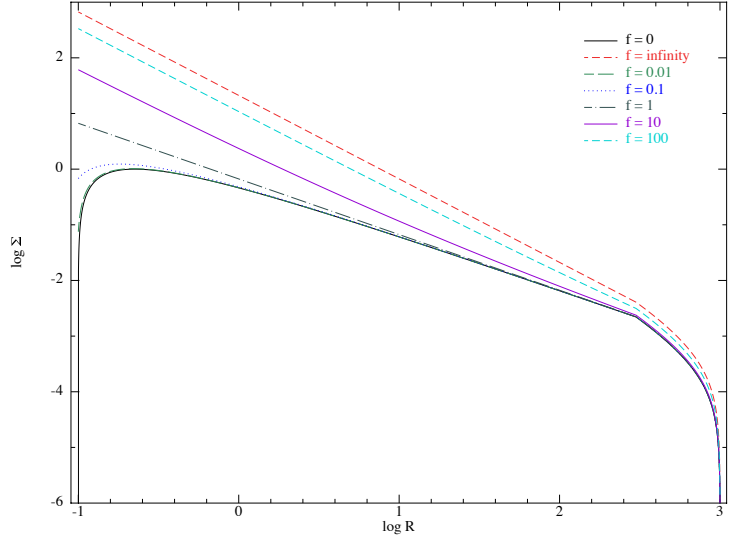}
  \caption{Steady disc structures with different $f$ values for an arbitrary input rate of mass ${\dot M}$. The plotted solutions are those in equations~\ref{s1} and \ref{s2}, and we have assumed arbitrarily that $\nu \propto R$. The results are scaled so that the peak surface density of the accretion disc solution ($f=0$) is unity. The disc inner edge is at $R_{\rm in} = 0.1$, the outer edge is at $R_{\rm out} = 1000$ and the mass is added at $R_{\rm add} = 300$. The outer disc boundary condition is zero torque, so the surface density goes to zero there. The inner disc boundary condition is determined by equation~\ref{innerBC} with the value of $f$ indicated in the legend. For $f=0$ (i.e. zero torque) the surface density goes to zero at the inner edge. For all other values of $f$ the surface density is non-zero at the inner edge. For $f \lesssim 0.1$ the solutions closely match the accretion disc solutions, while for $f \gtrsim 10$ the solutions more strongly resemble decretion disc solutions.}
  \label{fig1}
\end{figure}

To illustrate these steady solutions, we plot in Fig.~\ref{fig1} the surface density profile for discs with different $f$ values for the following parameters: $R_{\rm in} = 0.1$, $R_{\rm out} = 1000$, $R_{\rm add} = 300$, and $\nu = kR$ with $k=1$, and we have normalised the surface densities to the peak value for the accretion disc ($f=0$). This figure shows that when $f \lesssim 0.1$ the solutions closely resemble accretion disc solutions, and for $f \gtrsim 10$ the solutions more closely resemble decretion disc solutions.

\section{Time-dependent discs}   
\label{timedep}

We next consider the time-dependent behaviour of discs with non-zero inner torque. We consider the case in which the viscosity is a given function of radius. This keeps things simple as the evolution equation is then linear in $\Sigma$, but the basic physics is unaltered.

\subsection{Analytic solutions}
\label{timedepa}

For the case in which the viscosity is a power law in radius, the general solutions for both standard accretion ($f = 0$) and decretion ($f \rightarrow \infty$) discs are presented by \cite{Tanaka:2011aa}. \cite{Tanaka:2011aa} presents the Green's function (i.e. the solution corresponding to a disc which starts as a ring of mass at some radius $R_{\rm add}$ at time $t = 0$) for the case where the viscosity is a power law of radius, $\nu(R) \propto R^n$. Since in this case the problem for the evolution of surface density $\Sigma$ is a linear one, this provides the complete solution. Obtaining the Green's function involves setting up the radial behaviour of the solution in terms of integrals over modes, which involves Bessel functions, and then inverting the resulting integrals using Fourier-Bessel or Hankel transforms.

Pringle (1991) made use of the fact that for $\nu(R) \propto R$, the solution involves the Bessel functions $J_{1/2}(z) \propto \sin z\,/\!\sqrt{z}$ and $J_{-1/2}(z) \propto \cos z\,/\!\sqrt{z}$ which simplifies the algebra and makes the analysis more transparent. We employ this simplification below.

We assume that $\nu = kR$ for some constant $k$, we use as radial coordinate $x = \!\sqrt{R}$ and define a scaled surface density $\sigma(x,t) = \Sigma R^{3/2}$. We consider an infinite disc, with inner edge at $x=x_{\rm in}$. In this case, the inner boundary condition (\ref{innerBC}) becomes
\begin{equation}
\label{innerBC1}
\left[\sigma - f x \frac{\partial \sigma}{\partial x}\right]_{x_{\rm in}} = 0\,.
\end{equation}

The equation for the evolution of $\sigma$, obtained from (\ref{Sigma}), is a simple linear diffusion equation
\begin{equation}
\frac{\partial \sigma}{\partial t} = c^2 \frac{\partial^2 \sigma}{\partial x^2},
\end{equation}
where $c^2 = 3k/4$.

We look for the Green's function, which is the solution  of this equation with initial condition
\begin{equation}
\sigma(x,t=0) = \sigma_0 \delta(x - x_{\rm add}),
\end{equation}
where $x_{\rm add} > x_{\rm in}$. We note that once the Green's function has been obtained, it is then straightforward to obtain the solution for any combination of initial density distribution and time-dependent mass input distribution.

In Appendix A1, we show how the Green's function may be constructed for the case of a finite outer boundary $x_{\rm out}$. Here, and in Appendix A2 we consider the case of an infinite outer boundary, that is, the limit $x_{\rm out} \rightarrow \infty$.

For the decretion disc ($f \rightarrow \infty$), \cite{Pringle:1991aa} showed that the solution is then~\footnote{Note that there is a numerical error in the normalisation of the equivalent equation in \cite{Pringle:1991aa}.}
%\begin{eqnarray}
%\label{Greendec}
%\sigma(x,t) & = & \frac{\sigma_0}{2 c \sqrt{\pi t}} \\
%&& \times \left\{ \exp  \left[  - \frac{ (x - x_{\rm add})^2}{4c^2t} \right] +  \exp  \left[  - \frac{ (x + x_{\rm add} - 2x_{\rm in})^2}{4c^2t} \right] \right\} . \nonumber
%\end{eqnarray}  
\begin{equation}
\label{Greendec}
\sigma(x,t) = \frac{\sigma_0}{2c\sqrt{\pi t}}\times\left\{\exp\left[-\frac{(x-x_{\rm add})^2}{4c^2t}\right]+\exp\left[-\frac{(x+x_{\rm add}-2x_{\rm in})^2}{4c^2t}\right]\right\}\,.
\end{equation}  

For the accretion disc ($f = 0$), the corresponding solution can be obtained from the above by the Method of Images and is \citep[cf.][]{Tanaka:2011aa}
%\begin{eqnarray}
%\label{Greenacc}
%\sigma(x,t) & = & \frac{\sigma_0}{2 c \sqrt{\pi t}} \\
%&& \times \left\{ \exp  \left[  - \frac{ (x - x_{\rm add})^2}{4c^2 t} \right] -  \exp  \left[  - \frac{ (x + x_{\rm add} -2x_{\rm in})^2}{4c^2t} \right] \right\} . \nonumber
%\end{eqnarray}  
\begin{equation}
\label{Greenacc}
\sigma(x,t) = \frac{\sigma_0}{2c\sqrt{\pi t}}\times\left\{\exp\left[-\frac{(x-x_{\rm add})^2}{4c^2 t}\right]-\exp\left[-\frac{(x+x_{\rm add}-2x_{\rm in})^2}{4c^2t}\right]\right\}\,.
\end{equation}  

In \ref{appA} we demonstrate how the Green's function may be obtained in the case of the general finite torque boundary conditions (\ref{innerBC1}). The solution for $\sigma(x,t)$ in this case is
\begin{eqnarray}
\label{Greengeneral}
\sigma(x,t) & = & \frac{\sigma_0}{2c\sqrt{\pi t}}\times\left\{\exp\left[-\frac{(x-x_{\rm add})^2}{4c^2t}\right]+\exp\left[-\frac{(x+x_{\rm add}-2x_{\rm in})^2}{4c^2t}\right]\right\} \nonumber \\
&& -\,\frac{\sigma_0}{fx_{\rm in}}\exp\left(\frac{c^2t}{f^2x_{\rm in}^2}\right)\exp\left[\frac{x+x_{\rm add}-2x_{\rm in}}{fx_{\rm in}}\right]{\rm erfc}\left(\frac{c\sqrt{t}}{fx_{\rm in}}+\frac{x+x_{\rm add}-2x_{\rm in}}{2c\sqrt{t}}\right)\,.
\end{eqnarray}  
Here ${\rm erfc}(x)$ is the complimentary error function defined in (\ref{errorc}).

It is evident that in the limit $f \rightarrow \infty$ this expression agrees with the decretion disc Green's function (\ref{Greendec}). By using the asymptotic expansion for ${\rm erfc}(x \rightarrow \infty)$, which takes the form
\begin{equation}
{\rm erfc}(x \rightarrow \infty) \sim \frac{\exp(-x^2)}{x\sqrt{\pi}}\left[1+\mathcal{O}(x^{-2})\right]\,,
\end{equation}
it can be shown that (\ref{Greengeneral}) agrees with the (zero torque) accretion disc Green's function (\ref{Greenacc}) in the limit $f \rightarrow 0$.

\begin{figure}
  \begin{center}
  \includegraphics[width=0.4\columnwidth]{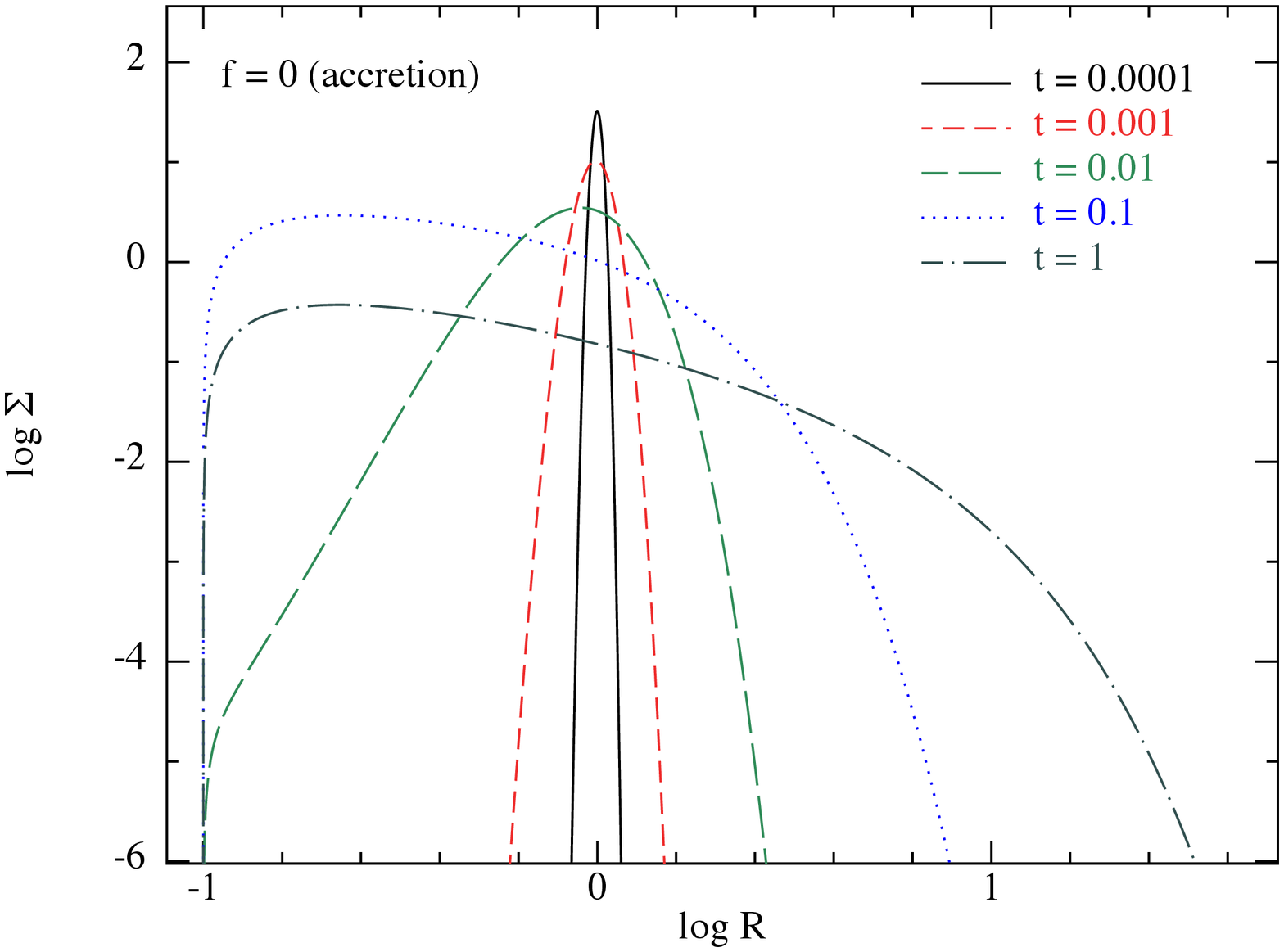}\hspace{0.5in}
  \includegraphics[width=0.4\columnwidth]{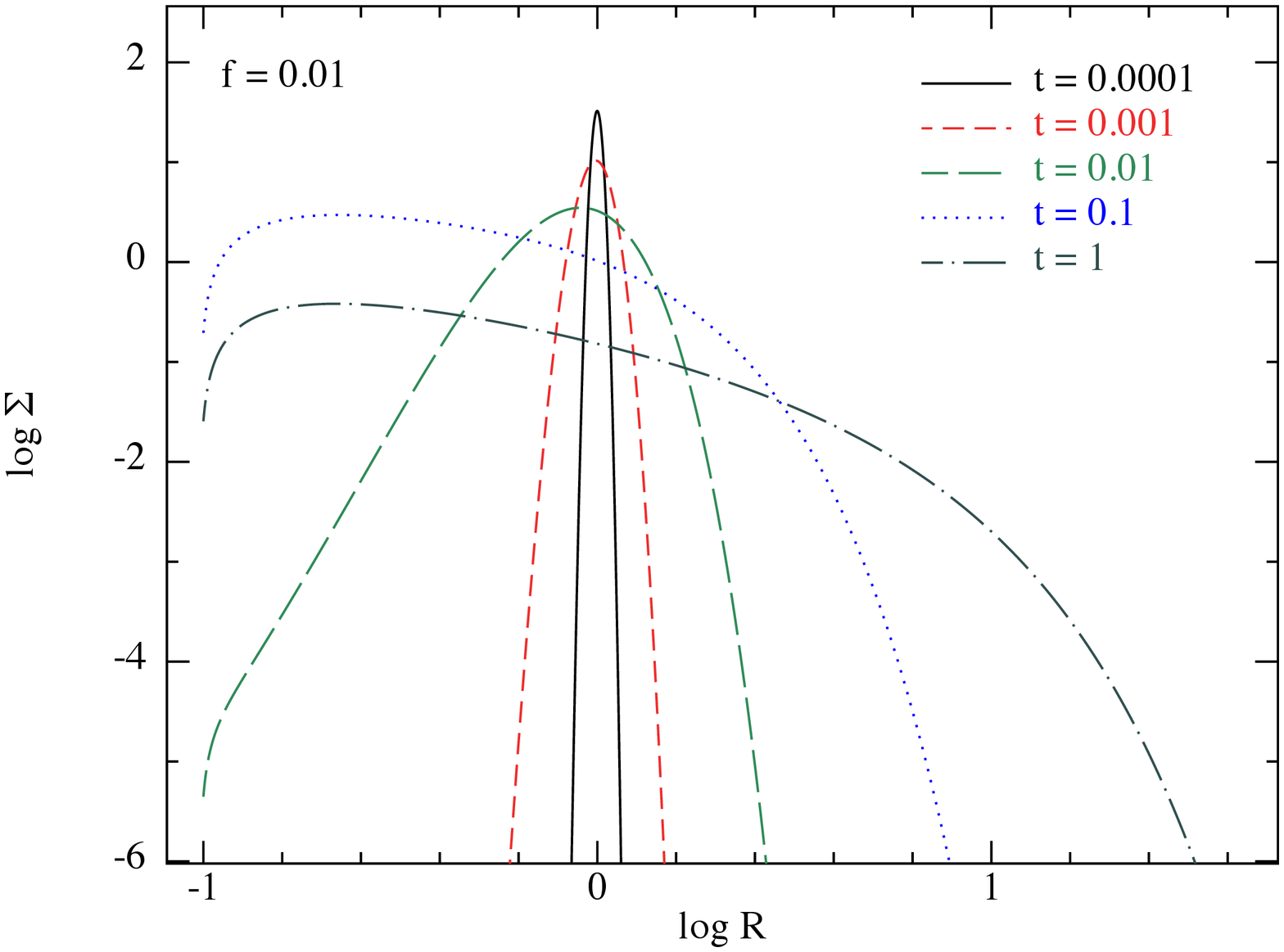}
  \includegraphics[width=0.4\columnwidth]{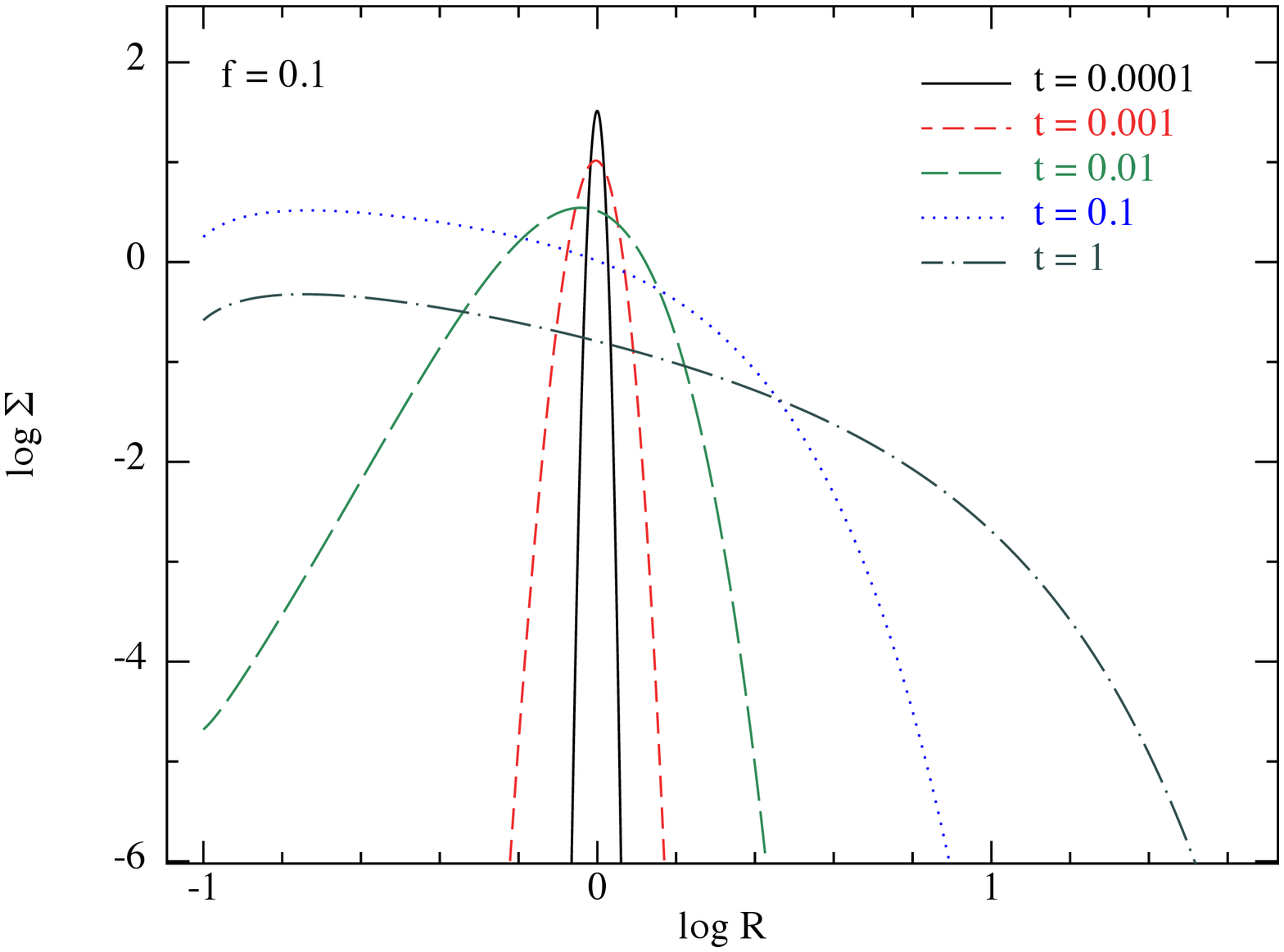}\hspace{0.5in}
  \includegraphics[width=0.4\columnwidth]{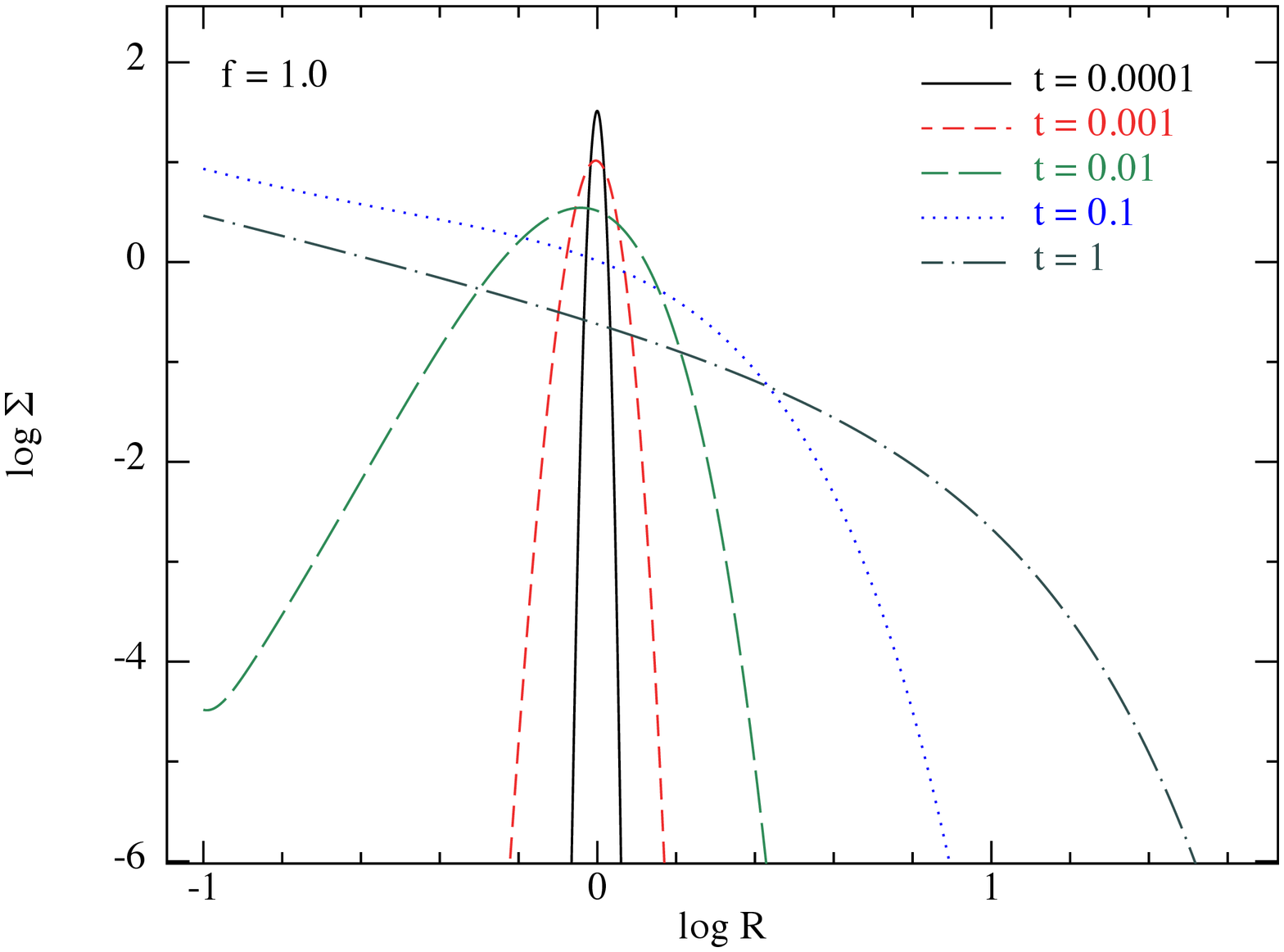}
  \includegraphics[width=0.4\columnwidth]{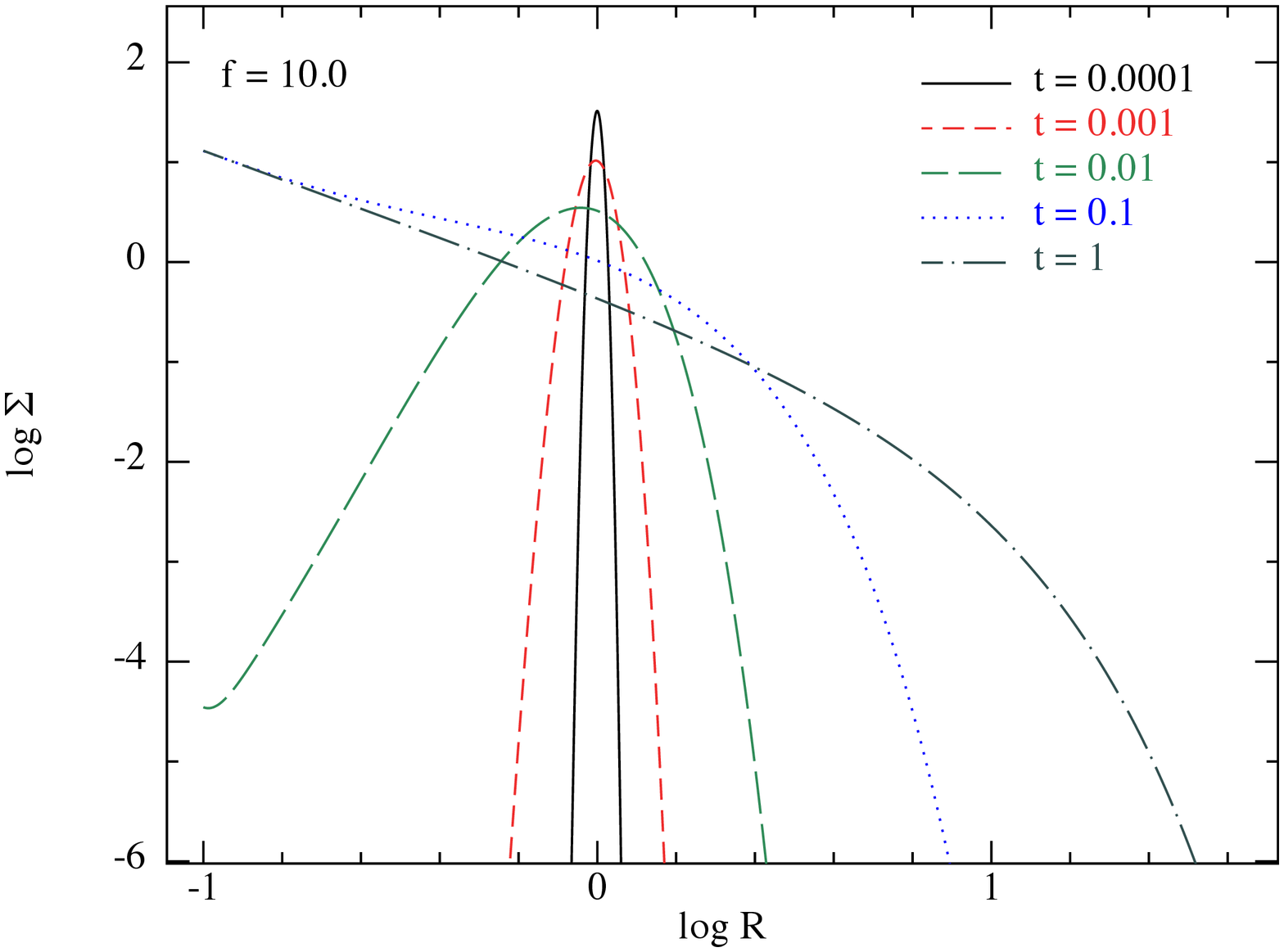}\hspace{0.5in}
  \includegraphics[width=0.4\columnwidth]{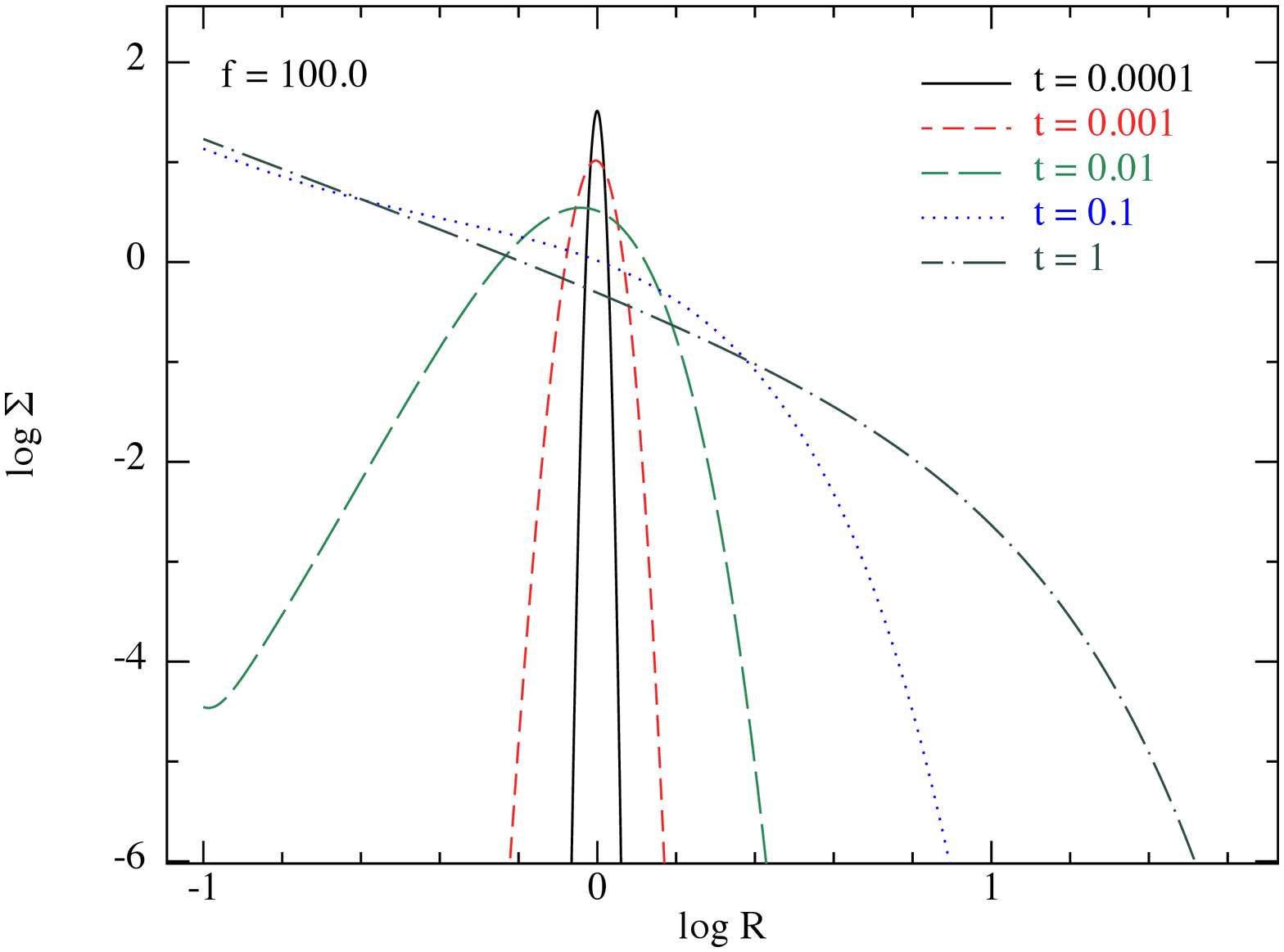}
  \includegraphics[width=0.4\columnwidth]{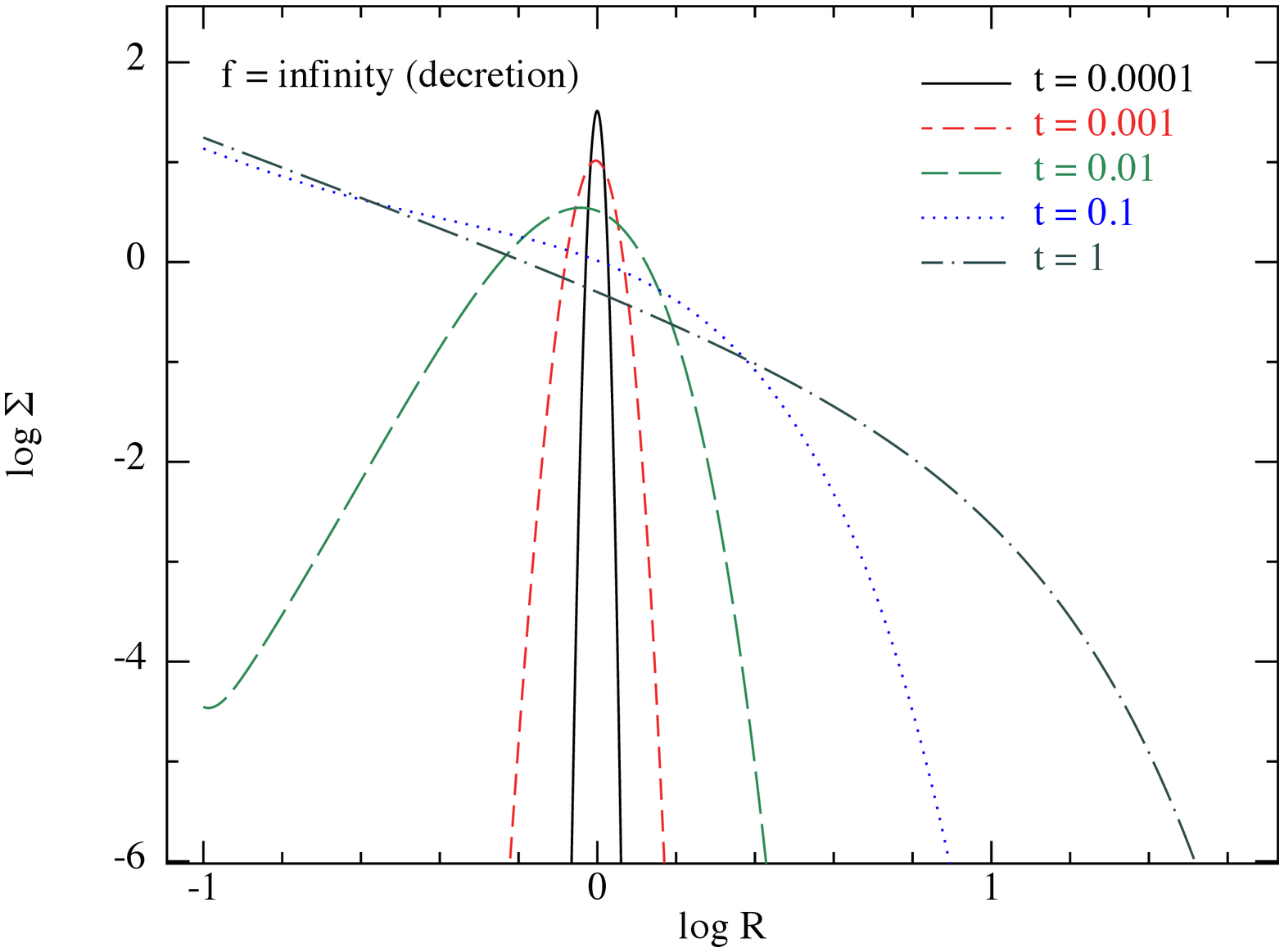}
  \end{center}
  \caption{The time-dependent disc solutions given by (\ref{Greengeneral}) for different values of $f$ as indicated in the top left corner of each panel. The different lines correspond to the times shown in the legend in units of $t_\nu = R_{\rm add}^2/\nu(R_{\rm add})$. The initial behaviour in each case is the same, and differences begin to appear once the disc has spread and begins to interact with the central boundary condition. For $f=0$, the surface density at $R_{\rm in}$ is zero, while for all other cases the value is non-zero. The disc outer boundary is at infinity, and thus in each case the discs spread outwards over time. For $f\lesssim 0.1$ the solutions are similar to the accretion disc ($f=0$) case. For $f\gtrsim 10$ the solutions are similar to the decretion ($f\rightarrow\infty$) case. The corresponding numerical solutions for this case (with an outer boundary at a finite radius) are plotted below in Fig.~\ref{fig4}.}
  \label{fig2}
\end{figure}

To depict the solutions we plot in Fig.~\ref{fig2} seven example cases with $f = 0$, $0.01$, $0.1$, $1$, $10$, $100$ and $\infty$. We plot $\Sigma = \sigma/R^{3/2}$ against $R = x^2$. For each case we plot the solution (from equations \ref{Greendec}, \ref{Greenacc} \& \ref{Greengeneral}) at different times\footnote{We note that for some combinations of parameters the exponential terms in (\ref{Greengeneral}) can become unwieldy. We provide an accurate expression at this point in \ref{estGreen}.} through the evolution corresponding to $t/t_\nu = 10^{-4}$, $10^{-3}$, $10^{-2}$, $10^{-1}$ and $1$, where $t_\nu = R_{\rm add}^2/\nu(R_{\rm add}) = R_{\rm add}/k$. For the plotted solutions we chose $k=1$, $\sigma_0 = 1$, $R_{\rm in} = 0.1$ and $R_{\rm add} = 1$. 

\begin{figure}
  \centering\includegraphics[width=0.5\columnwidth]{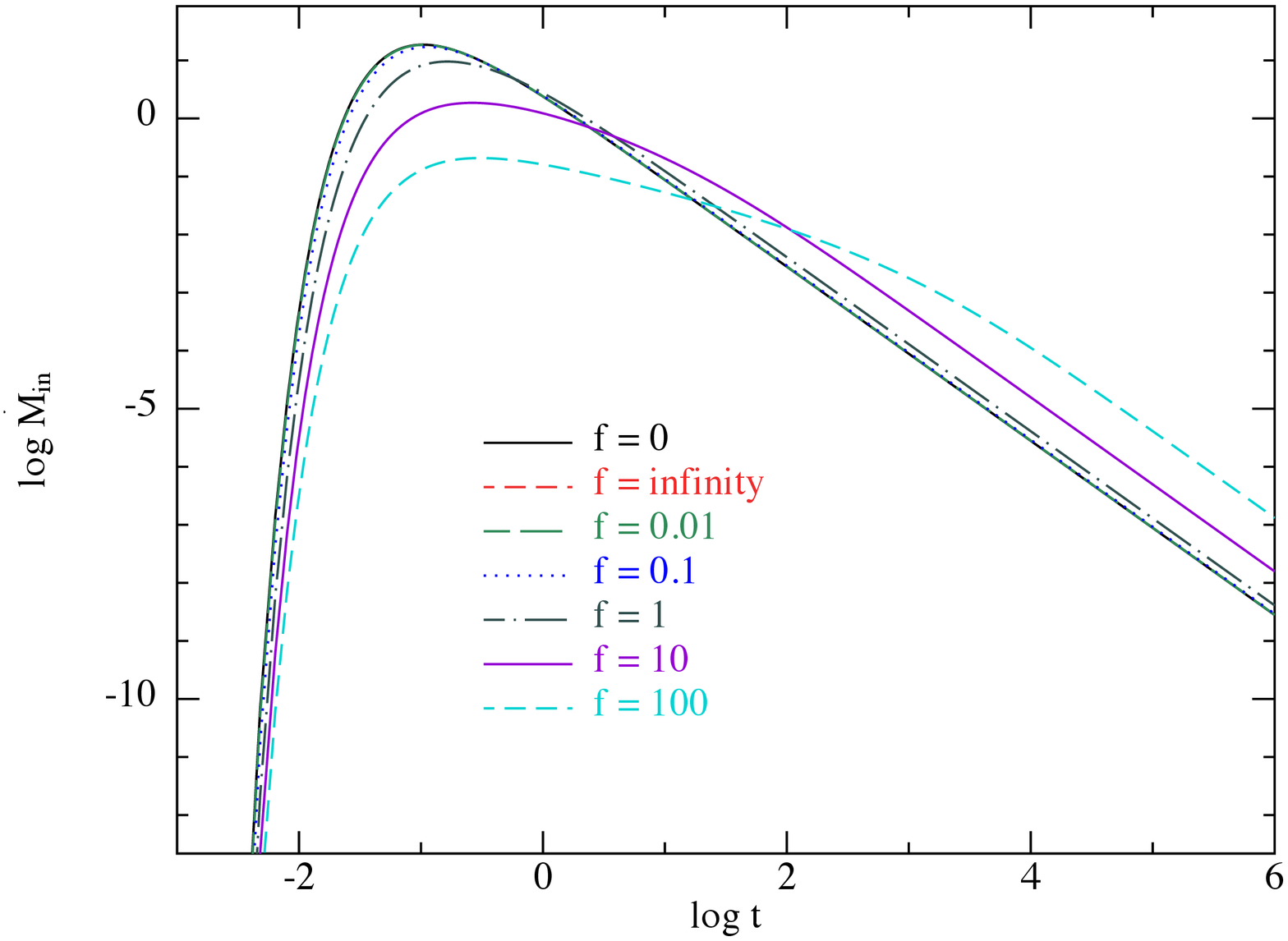}
  \centering\includegraphics[width=0.5\columnwidth]{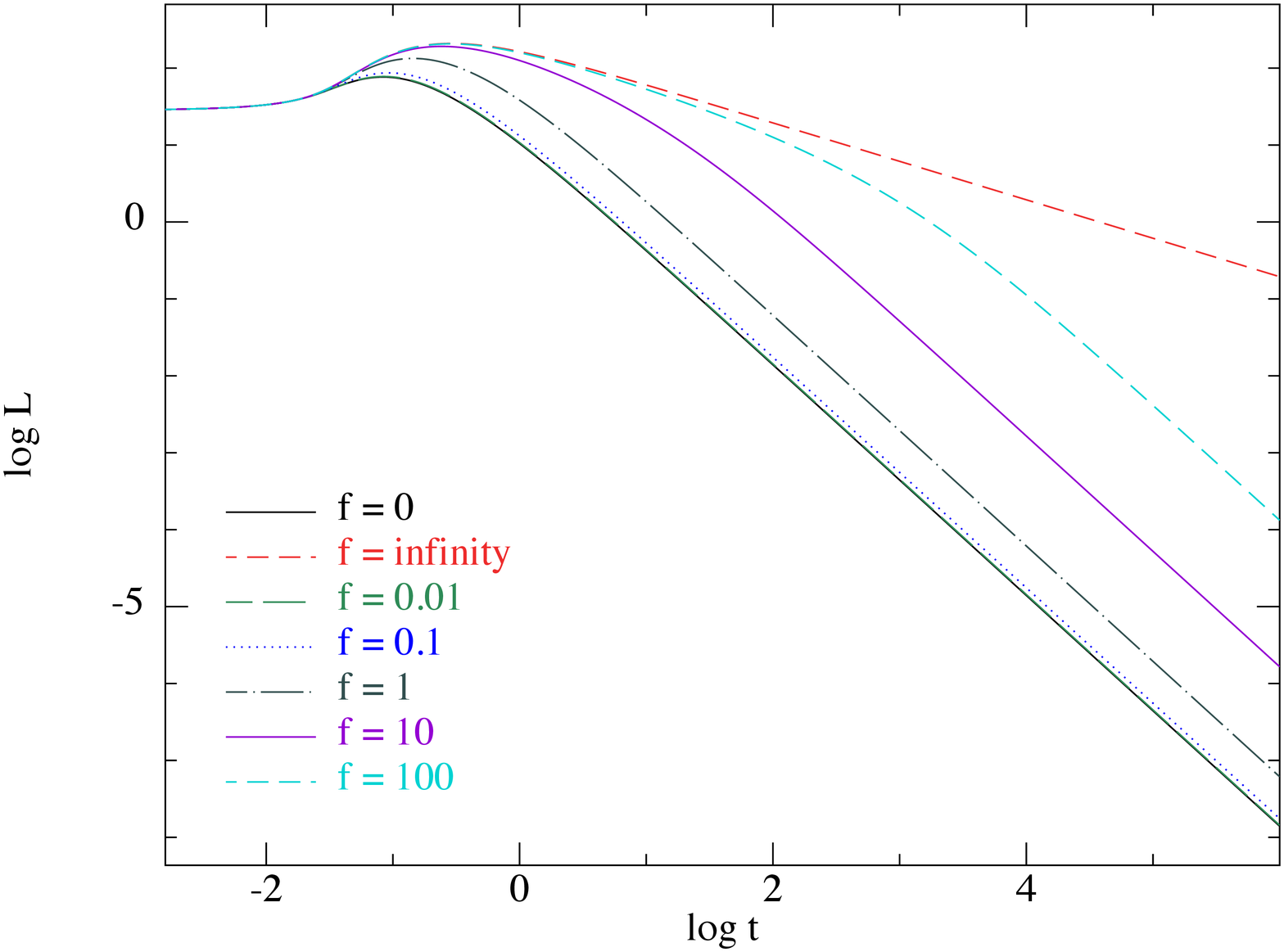}
  \centering\includegraphics[width=0.5\columnwidth]{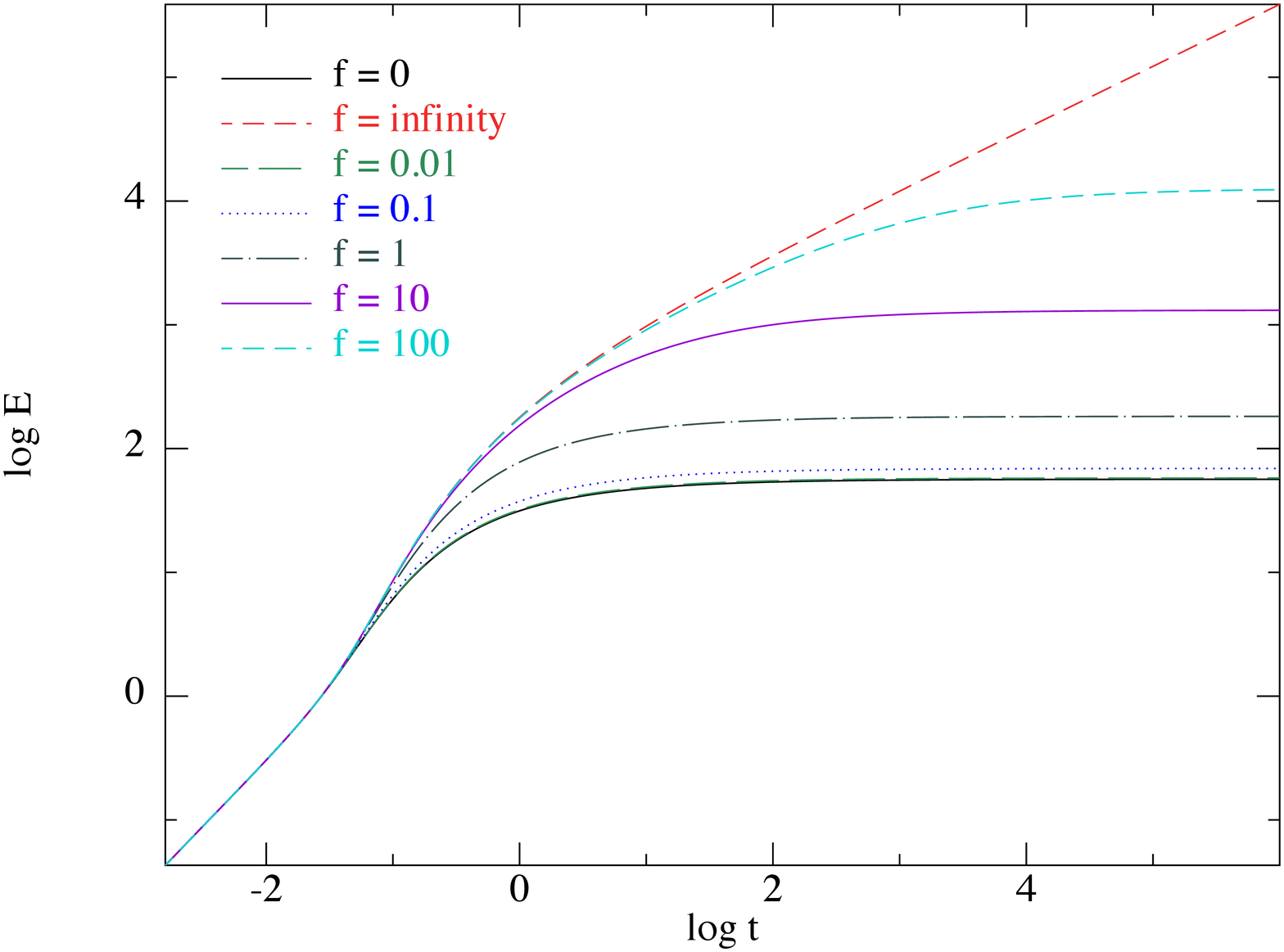}
  \caption{The accretion rate (top panel), luminosity (middle panel) and cumulative emitted energy (bottom panel) defined in equations (\ref{mdotin}), (\ref{Leq}) and (\ref{Eeq}) respectively are shown as functions of time. As for Figure~2, time is in units of the viscous timescale at $R  = R_{\rm add}$. The different cases with different $f$ values are shown with the line styles and colours indicated in the legend (note that for the ${\dot M}_{\rm in}$ plot the red line associated with the $f\rightarrow\infty$ case is not present as ${\dot M}_{\rm in} = 0$ in this case). In all of the panels the $f=0.01$ case (green dashed line) almost precisely overlaps the (f=0) case (black line). The $f=0.1$ case closely follows the $f=0$ case. Significant deviations are found for $f \gtrsim 1$. For large $f=10-100$ the solutions initially follow the decretion disc solution, before turning over to follow the accretion disc solution at late times (cf. eq.~\ref{Greengeneral}). From the figures it can be seen that this occurs at roughly $t\sim ft_\nu(R_{\rm add})$. For the decretion disc case the total emitted energy rises with time as energy is extracted from the inner boundary condition. For all finite values of $f$, the total energy emitted remains finite (cf. eq.~\ref{energy}), and at sufficiently late times all of the matter is accreted by the central accretor (cf. eq.~\ref{mdotin_steady}). The case $f=0$ shows the gravitational energy available from the initial ring. The additional energy emitted for $f>0$ comes from the torque applied at the inner boundary, and results in the late time emitted energy being increased by a factor $ \approx (1+2f)$.}
  \label{fig3}
\end{figure}

In Fig.~\ref{fig3} we show the central accretion rate (top panel), the disc luminosity (middle panel), and the total cumulative energy emitted by the disc (bottom panel) as functions of time for each of the $f$ values depicted in Fig.~\ref{fig2}. The central accretion rate is computed as
\begin{equation}
  {\dot M}_{\rm in} = \left[-2\pi R\Sigma \varv_R\right]_{\rm R_{\rm in}} = 6\pi R_{\rm in}^{1/2}\left.\frac{\partial S}{\partial R}\right|_{R_{\rm in}}\,,
\end{equation}
where in the last step we have assumed Keplerian rotation. Using the substitutions $x = R^{1/2}$, $\nu=kR$, and $\sigma=\Sigma R^{3/2}$ this becomes
\begin{equation}
  \label{mdotin}
  {\dot M}_{\rm in} = 3\pi k \left.\frac{\partial\sigma}{\partial x}\right|_{x_{\rm in}}\,.
\end{equation}
For a decretion disc we can see that $\partial\sigma/\partial x = 0$ at the inner boundary (cf. eq. \ref{Greendec}). For an accretion disc we can evaluate ${\dot M}_{\rm in}$ by differentiating (\ref{Greenacc}) and putting this into (\ref{mdotin}). While for a non-zero central torque disc we can make use of the boundary condition (\ref{innerBC1}) to obtain
\begin{equation}
  {\dot M}_{\rm in} = 3\pi k\frac{\sigma(x_{\rm in},t)}{fx_{\rm in}}\,.
\end{equation}

For the disc luminosity $L(t)$ in Fig.~\ref{fig3} we compute the integral
\begin{equation}
  \label{Leq}
  L(t) = \int_{R_{\rm in}}^{R_{\rm out}}2\pi R D(R,t)\,{\rm d}R\,
\end{equation}
where we have replaced the infinite outer disc radius with $R_{\rm out} = 1000$. This is sufficient as the dissipation rate per unit area, $D(R)$, falls off approximately with $\Omega^2$ at large radius.

We compute the total cumulative energy, $E$, emitted up to time $t$ through the integral
\begin{equation}
  \label{Eeq}
  E(t) = \int_{0}^{t}L(t^\prime){\rm d}t^\prime\,.
\end{equation}

In all of the Figures (2, 3) discussed above we find the general properties that for small $f< 0.1$ the results are essentially indistinguishable from the zero torque accretion disc case ($f = 0$), for moderate values of $f \approx 0.1$-$1$ the solutions closely resemble accretion disc solutions at all times with at most a small delay in the accretion of matter and a small factor increase in the total emitted energy, and for $f \gg 1$ the solutions exhibit significantly delayed accretion, a large increase in emitted energy and initially the solutions follow the decretion case before turning over at late times to follow the accretion solution (cf. \ref{Greengeneral}). For the case of a decretion disc ($f\rightarrow \infty$), a significant amount of energy and angular momentum is continually extracted from the inner boundary and thus the total emitted energy from the disc tends to infinity as time tends to infinity. For any finite value of $f$, the turnover to follow the accretion disc solution at late times implies that the total emitted energy remains finite and, for an outer disc boundary that is sufficiently far away (i.e. $R_{\rm out}\rightarrow \infty$), all of the matter is accreted on to the central object (cf. \ref{mdotin_steady}). For discs with $f > 0$ additional energy is supplied to the disc from the torque applied to the inner disc boundary by the central accretor, and this results in the total emitted energy being increased by a factor of $\approx (1+2f)$ (cf. \ref{energy}).

\citet[][Section 3.2]{Pringle:1991aa} gives that for an accretion disc ($f=0$) the accretion rate at late times scales as
\begin{equation}
  \label{latetime}
  {\dot M} \propto t^{-1-\lambda}\,,
\end{equation}
where $\lambda = (5m+4-2n)^{-1}$ and $\nu \propto \Sigma^m R^n$. We therefore also have that the late time luminosity $L\propto t^{-1-\lambda}$. Similarly for a decretion disc ($f\rightarrow \infty$), \cite{Pringle:1991aa} gives the time dependence of the late time central torque, and thus the time dependence of the late time luminosity as 
\begin{equation}
  {\dot M} \propto t^{-1+\lambda}\,,
\end{equation}
where in this case $\lambda = (4m+4-2n)^{-1}$\,.

For our choice of parameters here (Figs~\ref{fig2}, \ref{fig3}), we have $m=0$ and $n=1$, which implies that the late time accretion rate and luminosity for the accretion disc scales as $t^{-3/2}$ and the late time luminosity of the decretion disc scales as $t^{-1/2}$. It is apparent from Figure~3, that for all finite values of $f$ the late time behaviour resembles that of accretion discs, whereas for $f = \infty$ the late time behaviour resembles that of the decretion disc.

\subsection{Numerical solutions}
\label{numerical}
Above we have derived and shown the time-dependent solutions for discs with an inner boundary at $R_{\rm in}$ and an outer boundary at $R_{\rm out}\rightarrow\infty$. In this section we solve the time-dependent evolution equation (\ref{Sigma}) on a fixed (logarithmic) radial grid following the method outlined in, for example, \cite{Pringle:1991aa}. We have tested the number of grid points, using 250, 1000 and 4000, and find that the solutions are well-represented at all resolutions with increasing accuracy as resolution is increased. For the numerical simulations we focus on the spreading ring case  (Green's function) described in Section~\ref{timedepa}, but note that we have also performed simulations (not shown) with mass added to the grid over time and that these simulations find solutions which are essentially identical to those shown in Fig.~\ref{fig1}. We perform our calculations with the same parameters as the analytical solutions shown in Figs~\ref{fig2} \& \ref{fig3}, that is: $R_{\rm in} = 0.1$, $\nu = kR$ with $k=1$, $\sigma_0 = 1$ and $R_{\rm add} = 1$, and later we provide additional simulations with $\nu = k R^{3/2}$ to demonstrate a different viscosity\footnote{In principle, we could provide analytical solutions for more complex cases, but we consider this effort not worthwhile as the numerical solutions are highly accurate and inexpensive.}.

For the numerical case we must impose an outer boundary at a finite radius, and we choose $R_{\rm out} = 1000$. It is worth noting that at times $t\gtrsim t_\nu(R_{\rm out}) = R_{\rm out}/k$ the solutions will begin to diverge from the analytical solutions as the disc is then aware of the different outer boundary condition. For the outer boundary we apply a zero torque boundary condition. We run the numerical simulations to a time of $t = 1000t_\nu(R_{\rm add})$ (which in our units corresponds to $t=1000$ as $t_\nu(R_{\rm add}) = R_{\rm add}/k = 1$).

For the inner boundary condition we use a ghost grid point (labelled $i=0$) which exists at the logarithmically spaced point inside the inner disc edge (at $R_{\rm in}$, labelled $i=1$). The quantity of interest for the numerical integration of (\ref{Sigma}) is $S$, and in this case we have $S = \nu\Sigma R^{1/2}$. To define the value of $S$ at the boundary, i.e. $S_0$, we apply the boundary condition (\ref{innerBC}). Approximating the derivative at the inner boundary yields
\begin{equation}
  S_1 - 2fR_1\frac{S_1-S_0}{\Delta R} = 0\,,
\end{equation}
where $\Delta R = R_1-R_0$, and thus the value of $S_0$ is (for $f \ne 0$) given by
\begin{equation}
  S_0 = S_1\left(1 - \frac{1}{2f}\frac{\Delta R}{R_1}\right)\,.
\end{equation}
We note that for the accretion disc case of $f=0$ the grid point at $i=0$ is not required and one may simply set the surface density to zero at the inner boundary at the end of each timestep, while for decretion discs ($f\rightarrow\infty$) we have $S_0 = S_1$ which enforces no mass flow across the inner boundary. We note that as long as the inner disc regions are sufficiently well resolved (given by $\Delta R/R_1 \ll 2f$), then the value of $S_0$ is positive and the numerical scheme conserves mass to machine precision (angular momentum is also conserved to machine precision when one accounts for the torque applied by the boundary condition).

\begin{figure}
  \begin{center}
  \includegraphics[width=0.4\columnwidth]{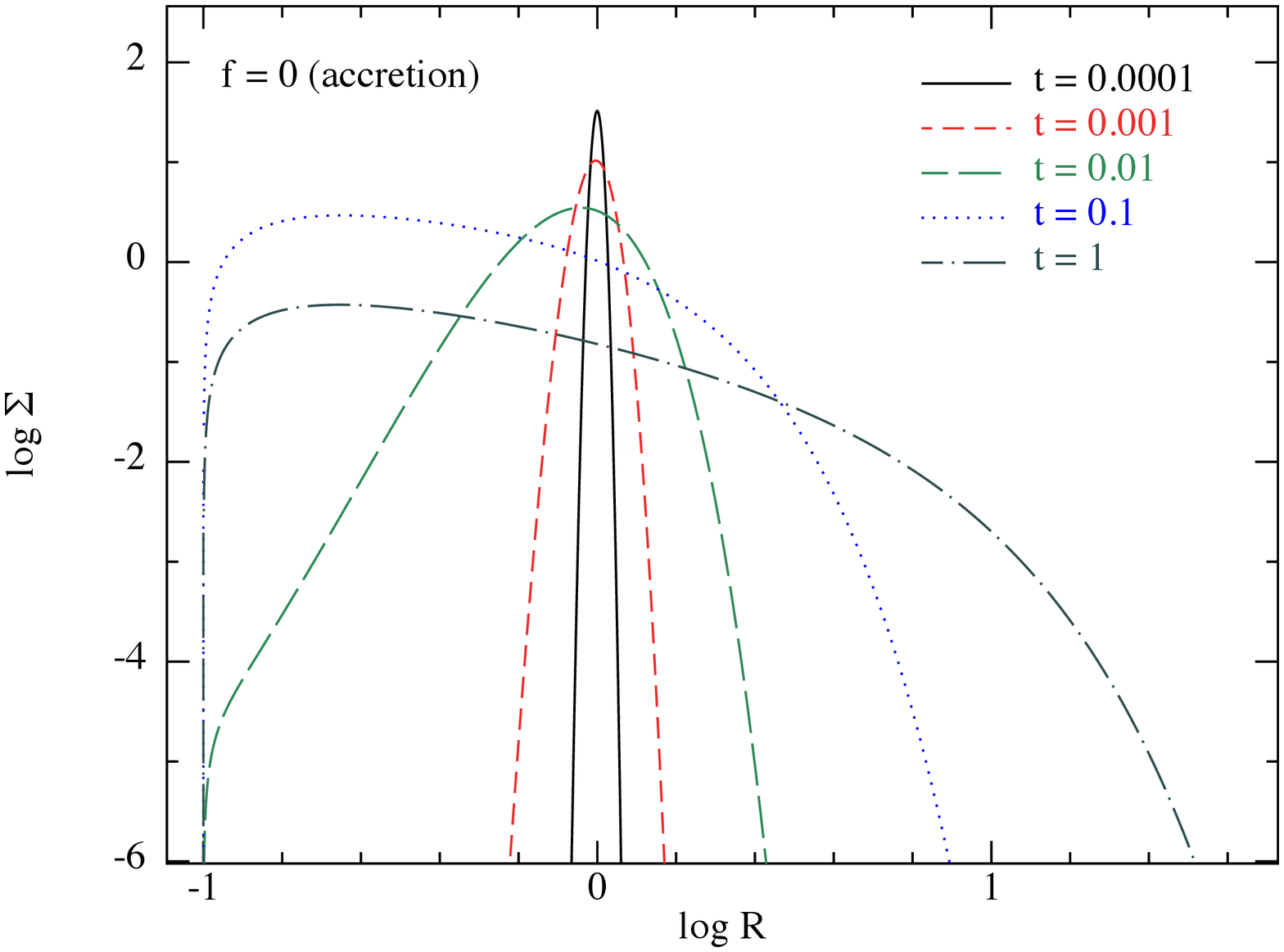}\hspace{0.5in}
  \includegraphics[width=0.4\columnwidth]{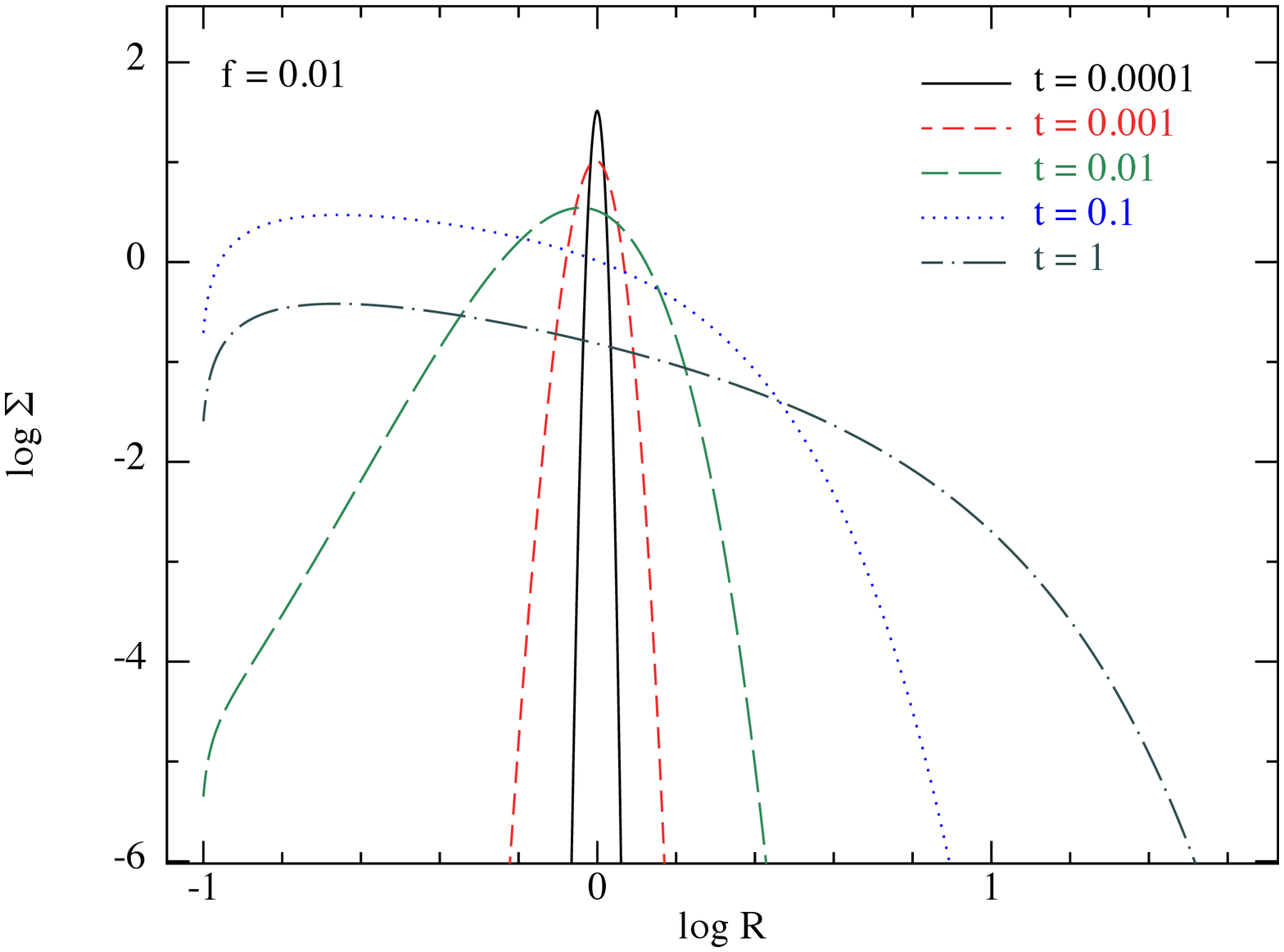}
  \includegraphics[width=0.4\columnwidth]{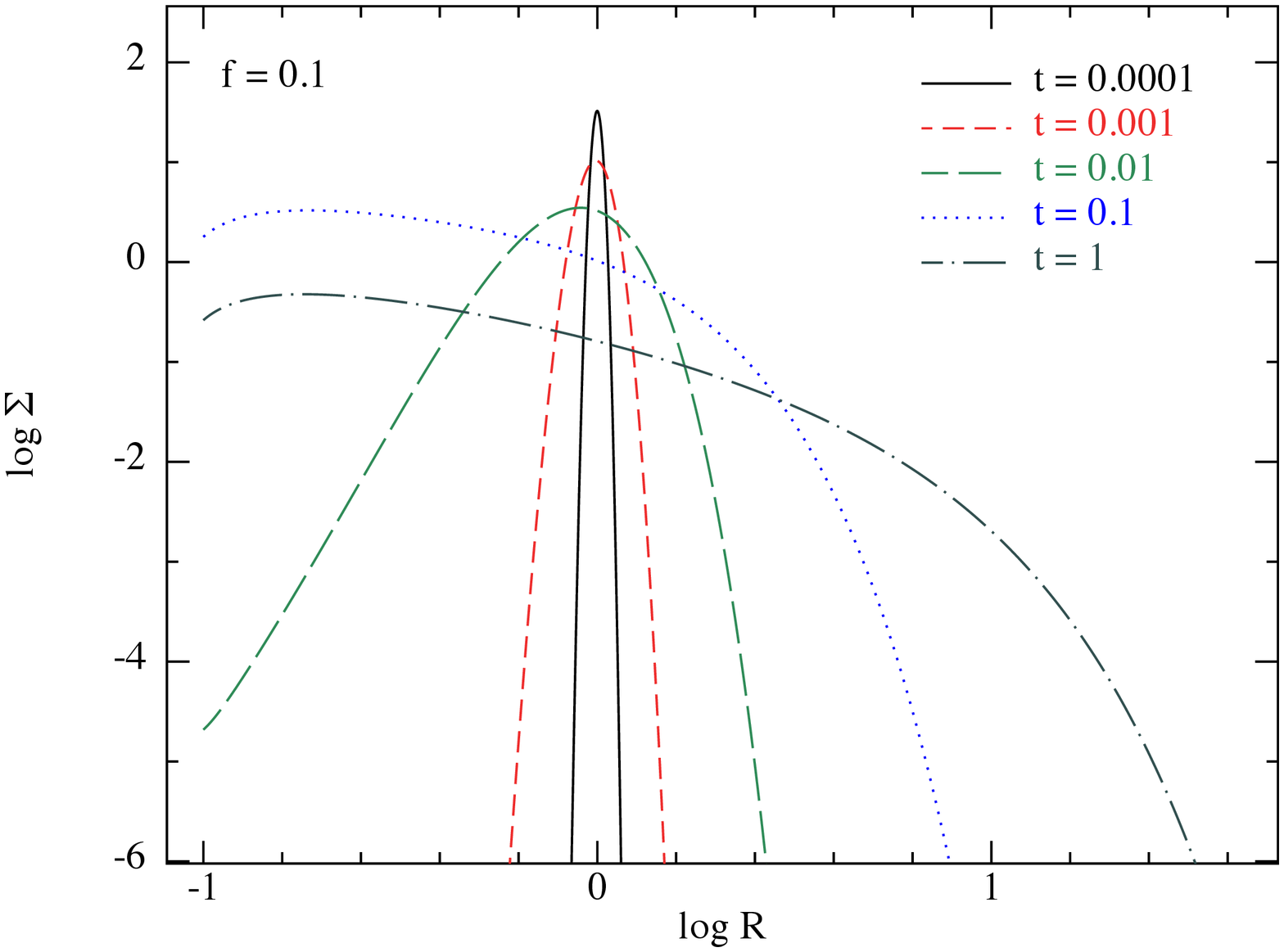}\hspace{0.5in}
  \includegraphics[width=0.4\columnwidth]{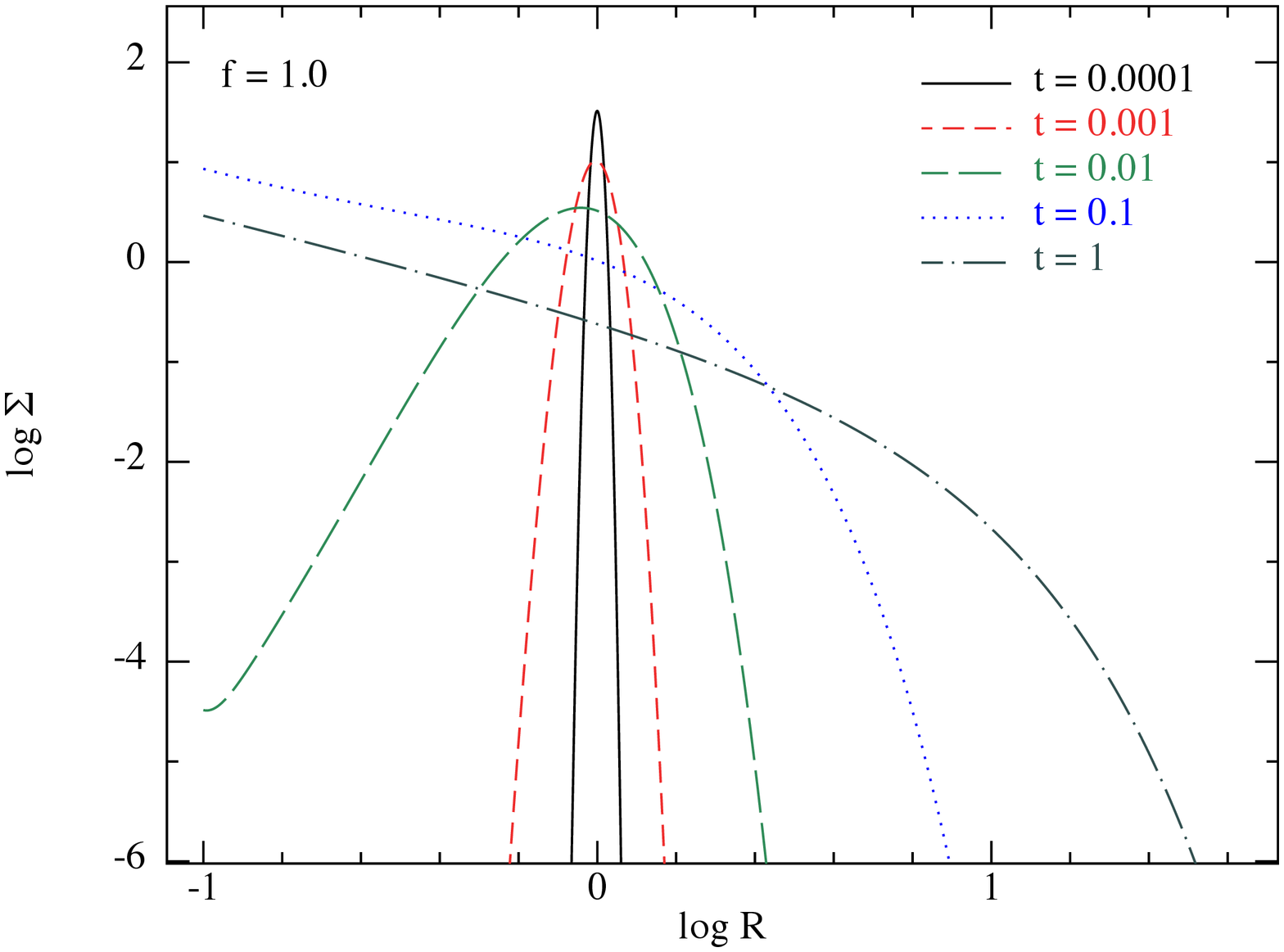}
  \includegraphics[width=0.4\columnwidth]{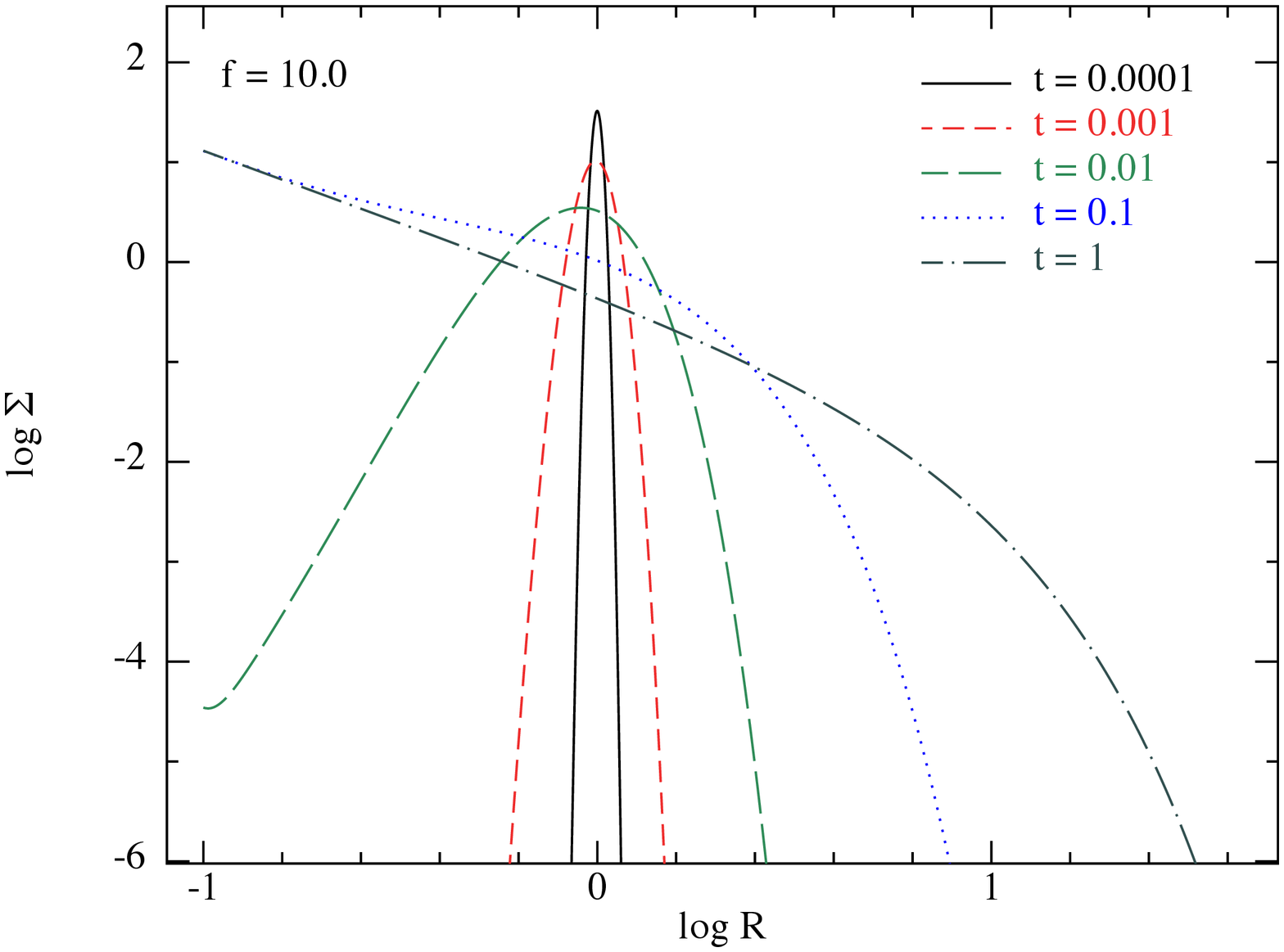}\hspace{0.5in}
  \includegraphics[width=0.4\columnwidth]{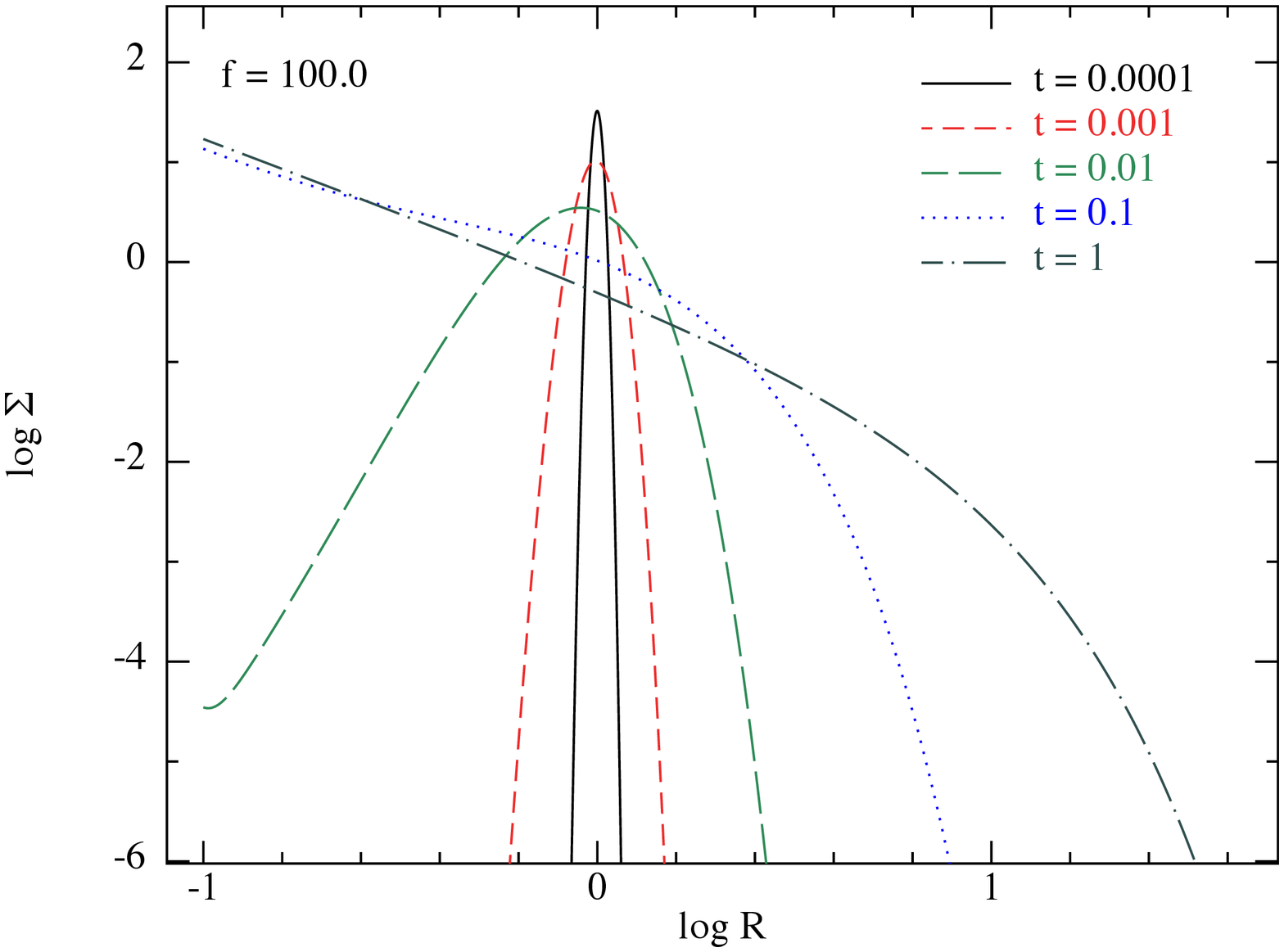}
  \includegraphics[width=0.4\columnwidth]{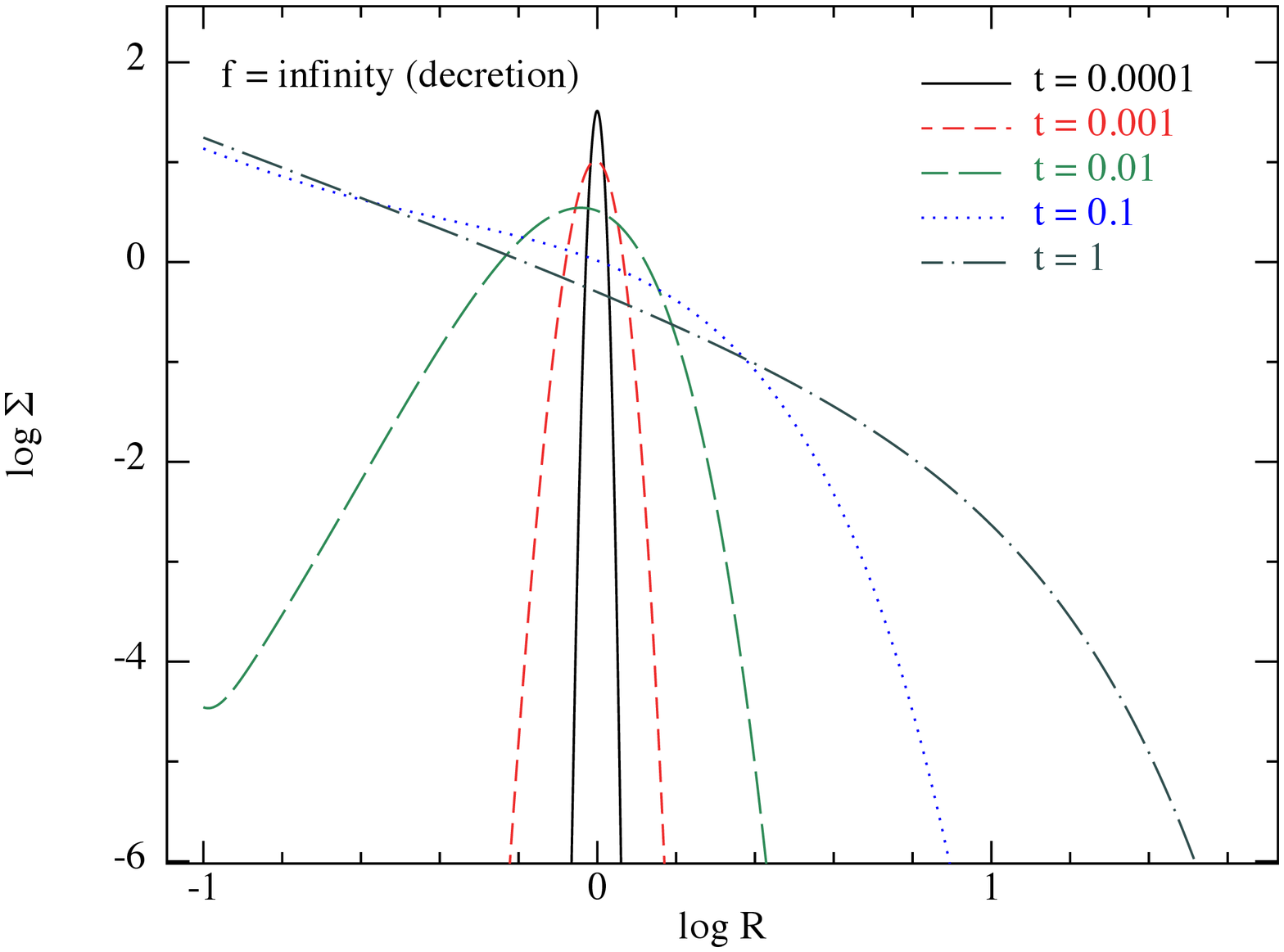}
  \end{center}
  \caption{Numerical solutions to the disc equations (described in Section~\ref{numerical}) for the case depicted in Fig.~\ref{fig2}, i.e. with $\nu = kR$. For the numerical case shown here we impose a zero torque outer boundary at $R_{\rm out}=1000$. Thus at times $t \ll t_\nu(R_{\rm out})$ we expect similar results, but once $t \gtrsim t_\nu(R_{\rm out})$ we expect the solutions to diverge as matter is lost through the outer boundary in this case. Here we have $k=1$ and thus $t_\nu(R_{\rm out}) = 1000$. We can see that, by comparing this figure with Fig.~\ref{fig2} the analytical and numerical solutions are in close agreement for all values of $f$ that we have considered.}
  \label{fig4}
\end{figure}

\begin{figure}
  \centering\includegraphics[width=0.5\columnwidth]{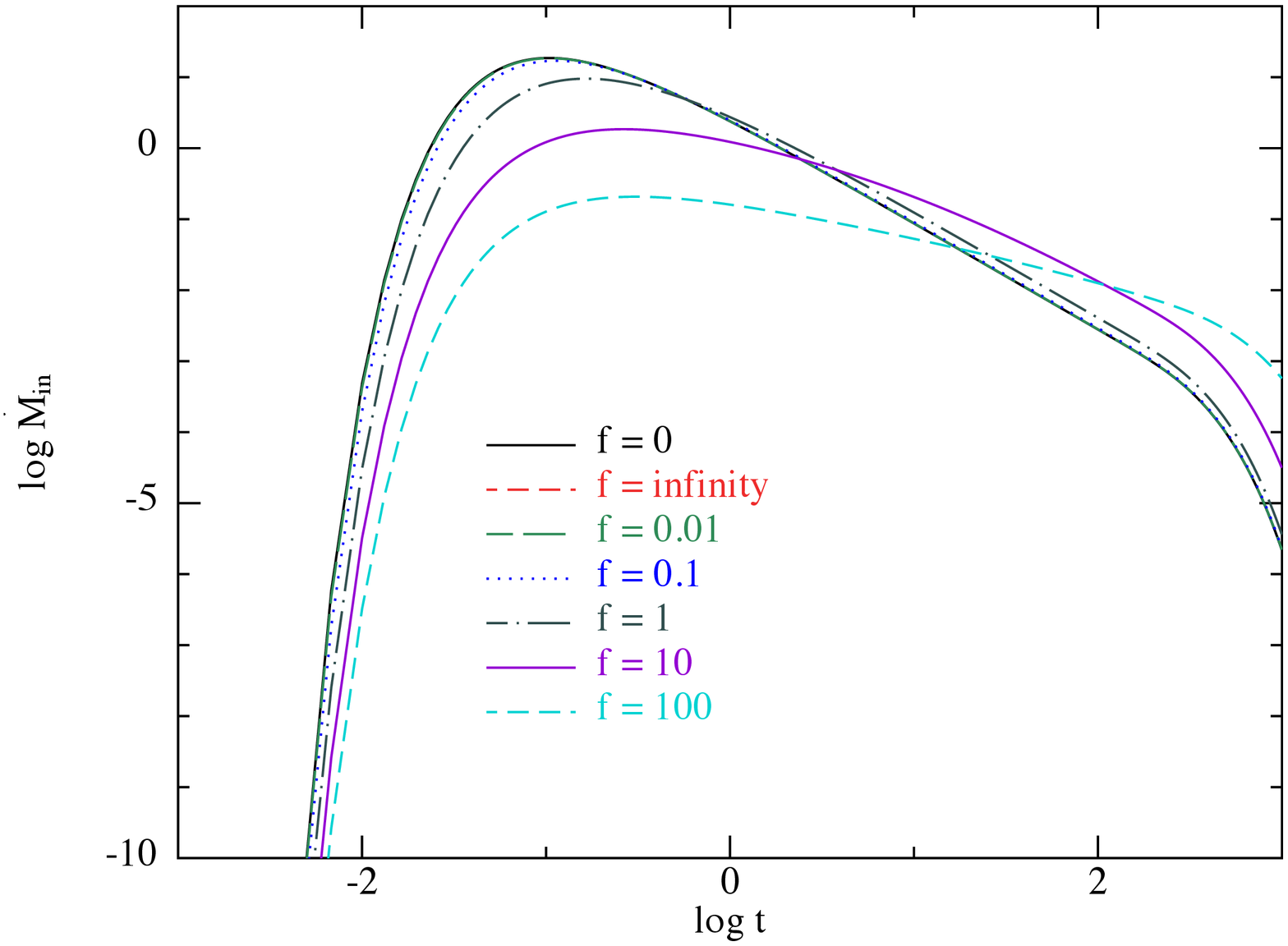}
  \centering\includegraphics[width=0.5\columnwidth]{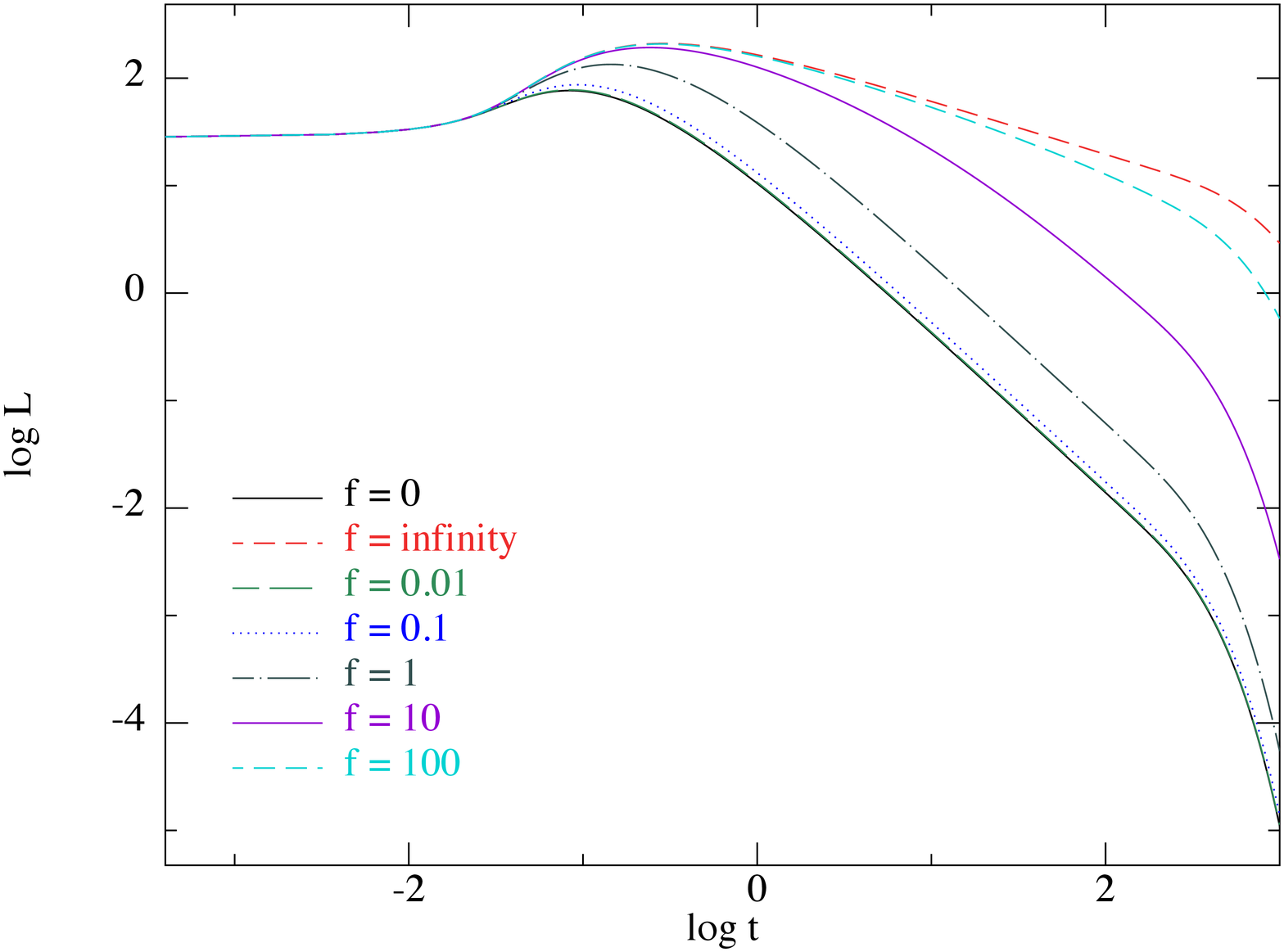}
  \centering\includegraphics[width=0.5\columnwidth]{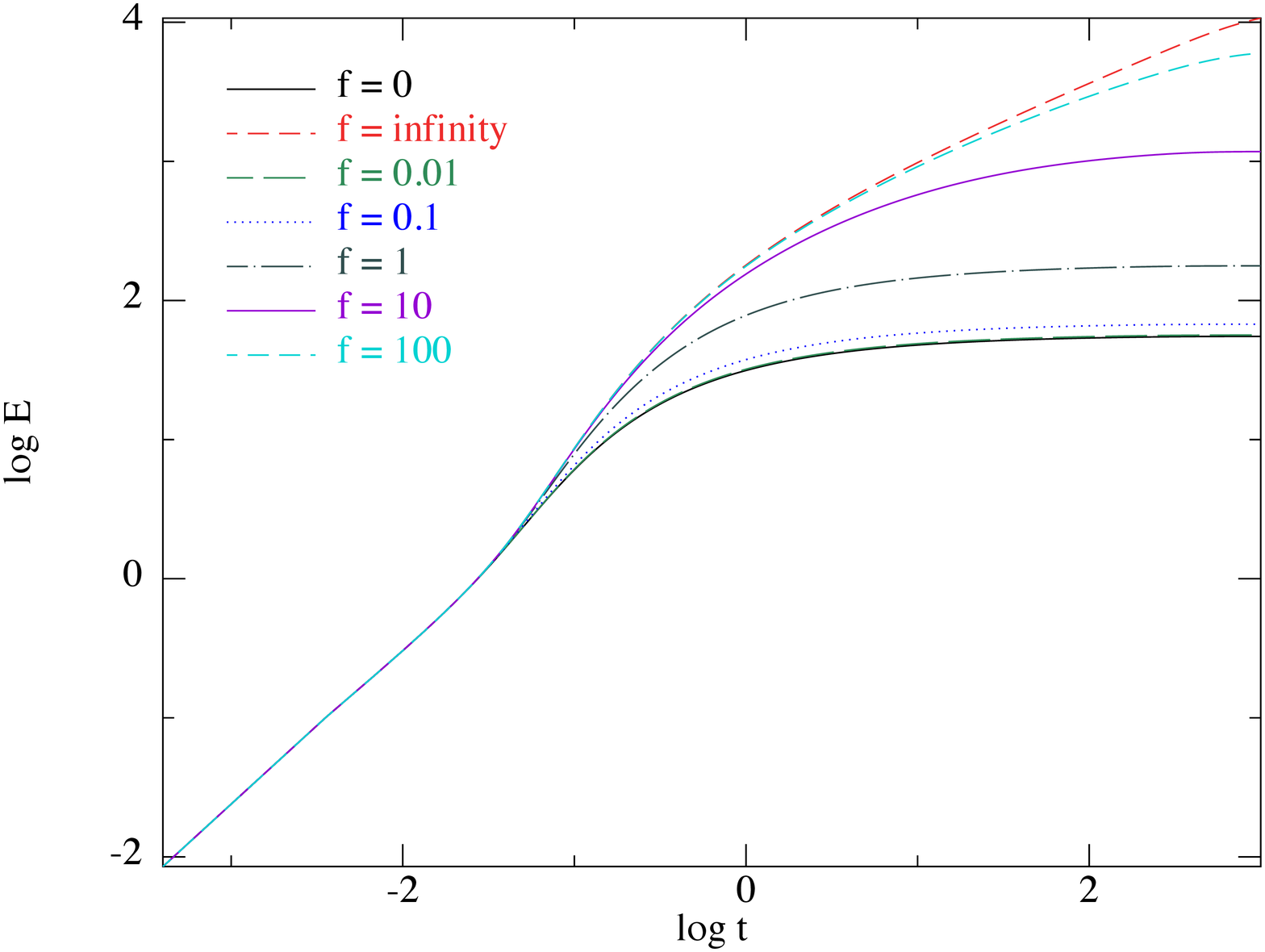}
  \caption{The accretion rate (top panel), luminosity (middle panel) and cumulative emitted energy (bottom panel) defined in equations (\ref{mdotin}), (\ref{Leq}) and (\ref{Eeq}) respectively, calculated from the numerical solutions to the disc equations (depicted in Fig.~\ref{fig4}). For this numerical case, the solutions diverge at late times from the infinite disc case as we impose a zero torque outer boundary at $R_{\rm out} = 1000$. This means that for times $t\gtrsim t_\nu(R_{\rm out}) = 1000$ the disc mass begins to decrease (as matter is lost through the outer boundary), and thus the central accretion rate and disc luminosity both begin to decline sharply with time as seen in this figure when compared to Fig~\ref{fig3}. In Appendix A1 we show that for a finite outer boundary, at times $t \gtrsim t_\nu(R_{\rm out})$ the solution decreases exponentially on a timescale $t_0 \approx (4/3 Y_1)^2 t_\nu(R_{\rm out})$ where $\frac{1}{2} \pi < Y_1 < \pi$.}
  \label{fig5}
\end{figure}

The results from the numerical simulations are presented in Figs.~\ref{fig4} \& \ref{fig5}, and these can be directly compared to Figs.~\ref{fig2} \& \ref{fig3}. As can be seen the numerical solutions provide close agreement with the analytical solutions. At late times, $t \gtrsim t_\nu(R_{\rm out})$ the numerical solutions diverge from the analytical solutions as significant amounts of matter have reached the disc outer edge by this point and are lost through the outer (zero torque) boundary. Closer agreement at these times can be achieved simply by extending the numerical grid to larger radii. In Appendix A1 we show that for a finite outer boundary, at times $t \gtrsim t_\nu(R_{\rm out})$ the solution decreases exponentially on a timescale $t_0 \approx (4/3 Y_1)^2 t_\nu(R_{\rm out})$ where $\frac{1}{2} \pi < Y_1 < \pi$, and this is seen in Fig.~\ref{fig5}.

\begin{figure}
  \begin{center}
  \includegraphics[width=0.4\columnwidth]{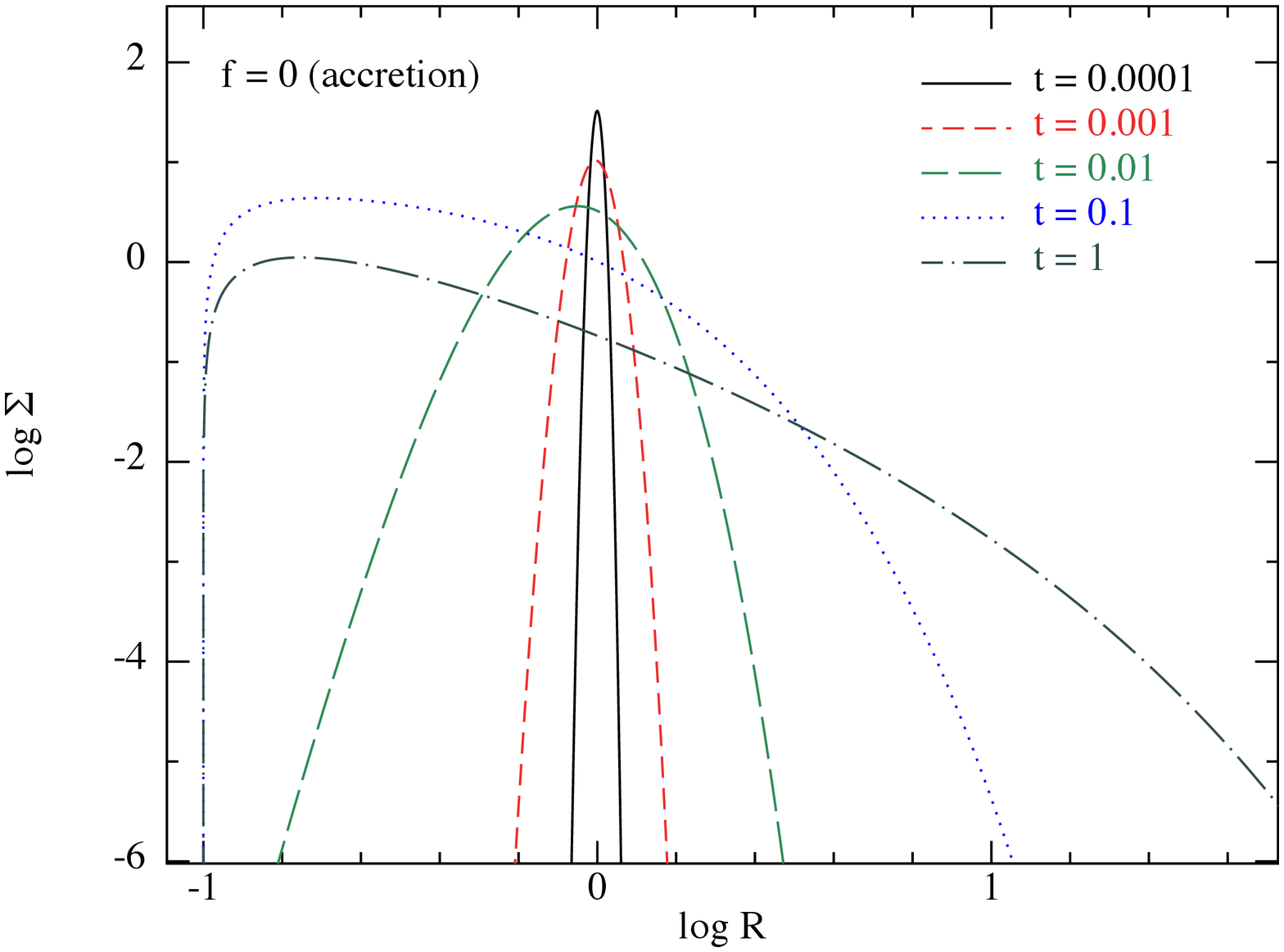}\hspace{0.5in}
  \includegraphics[width=0.4\columnwidth]{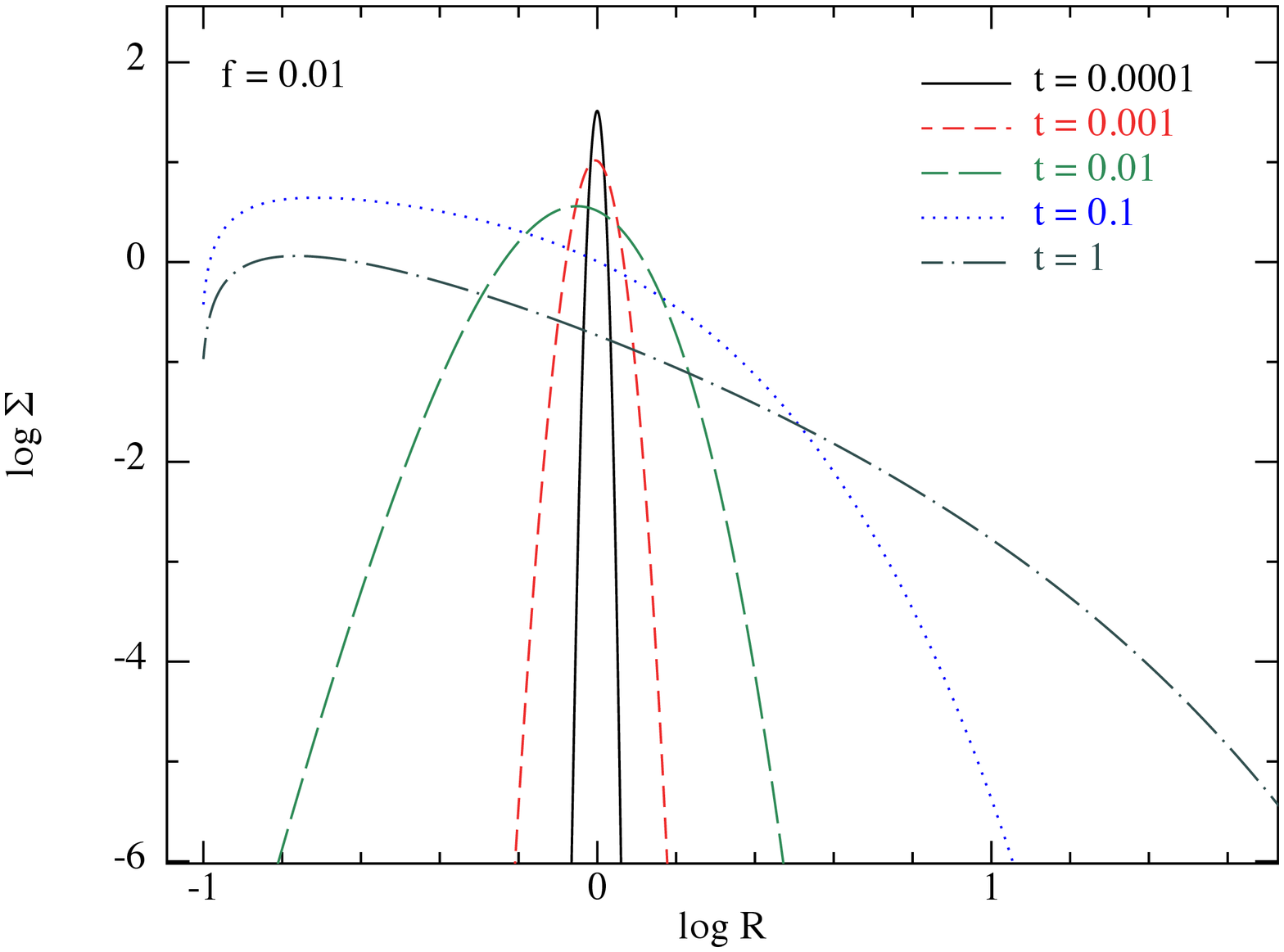}
  \includegraphics[width=0.4\columnwidth]{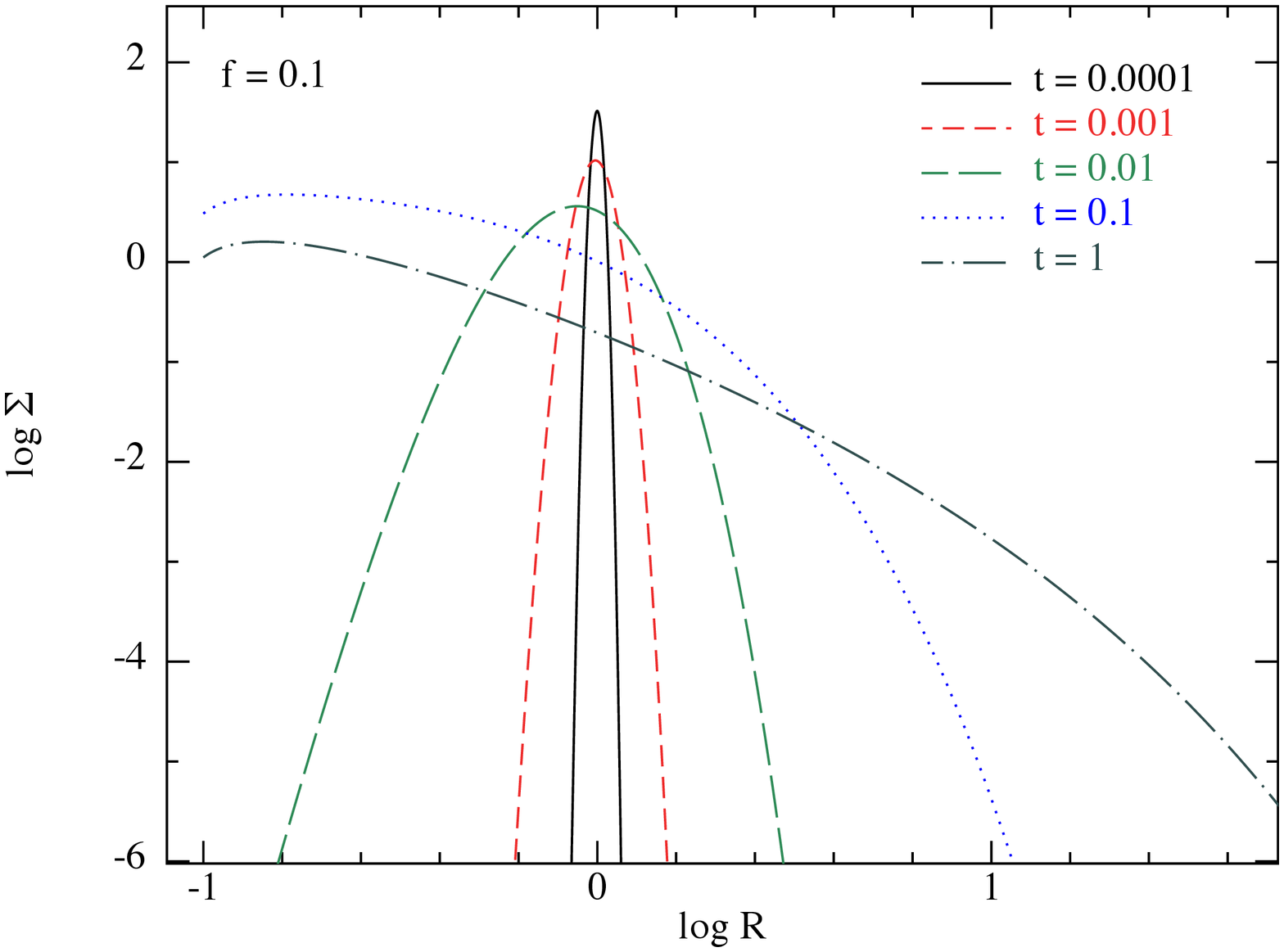}\hspace{0.5in}
  \includegraphics[width=0.4\columnwidth]{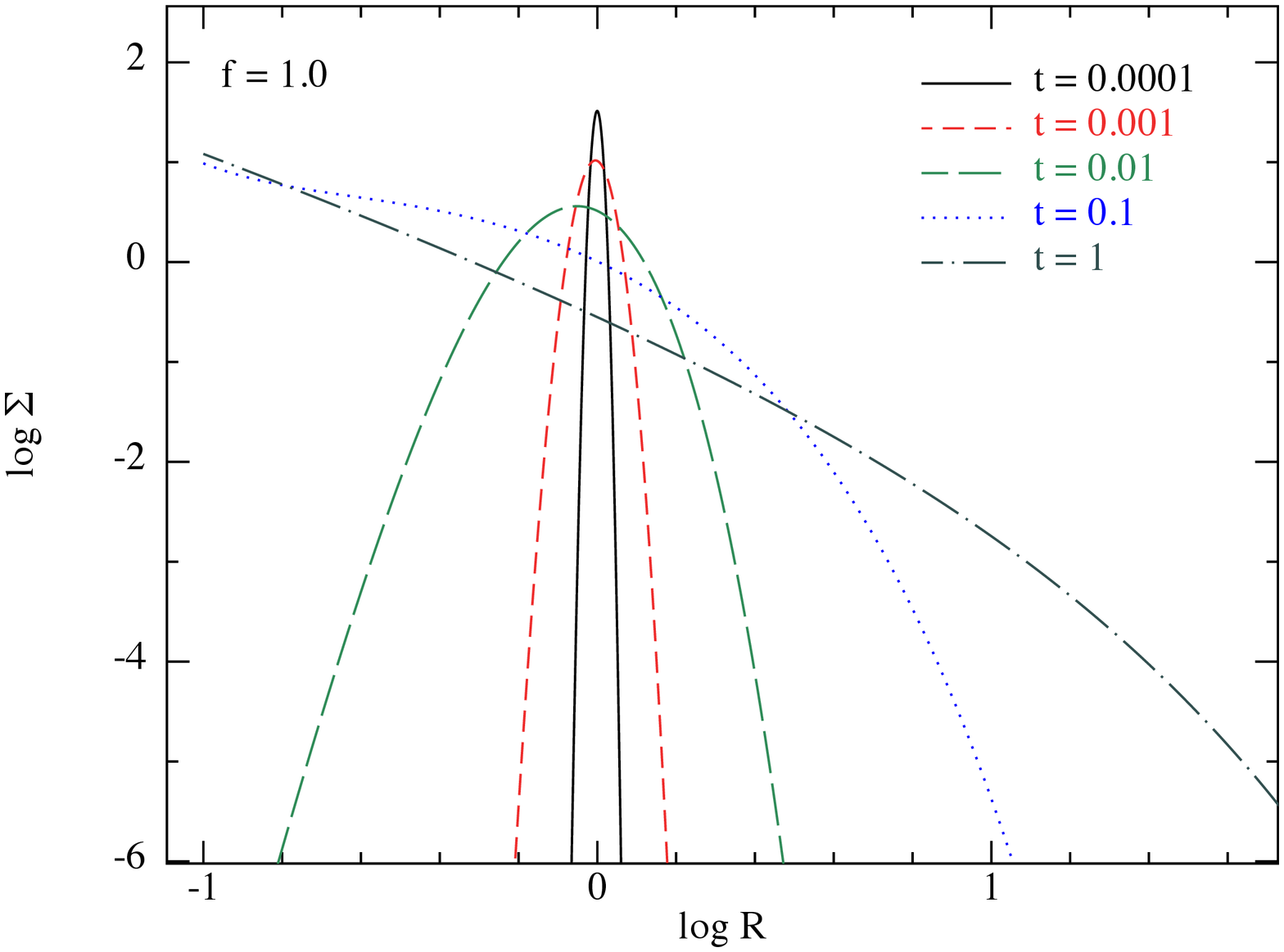}
  \includegraphics[width=0.4\columnwidth]{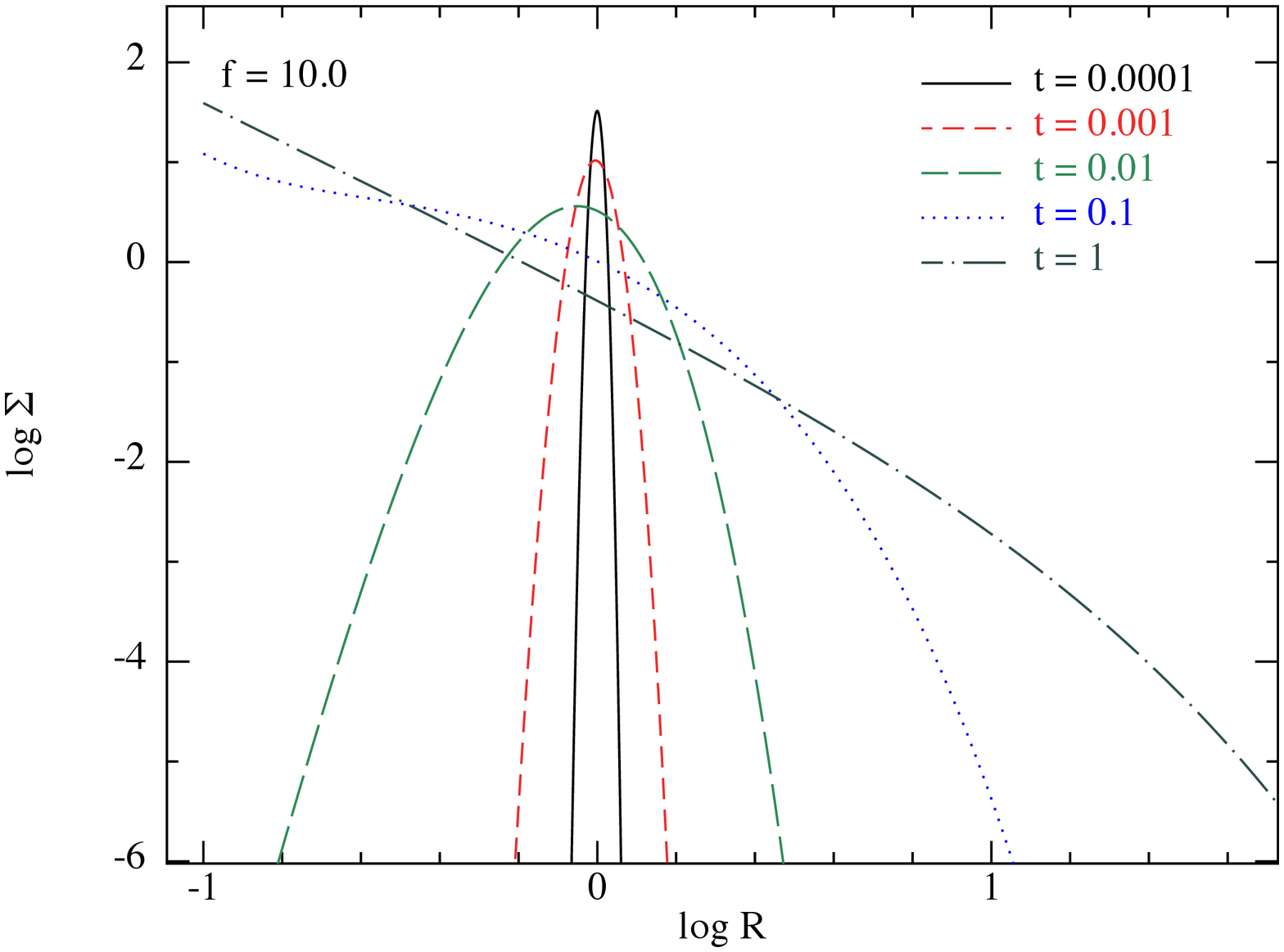}\hspace{0.5in}
  \includegraphics[width=0.4\columnwidth]{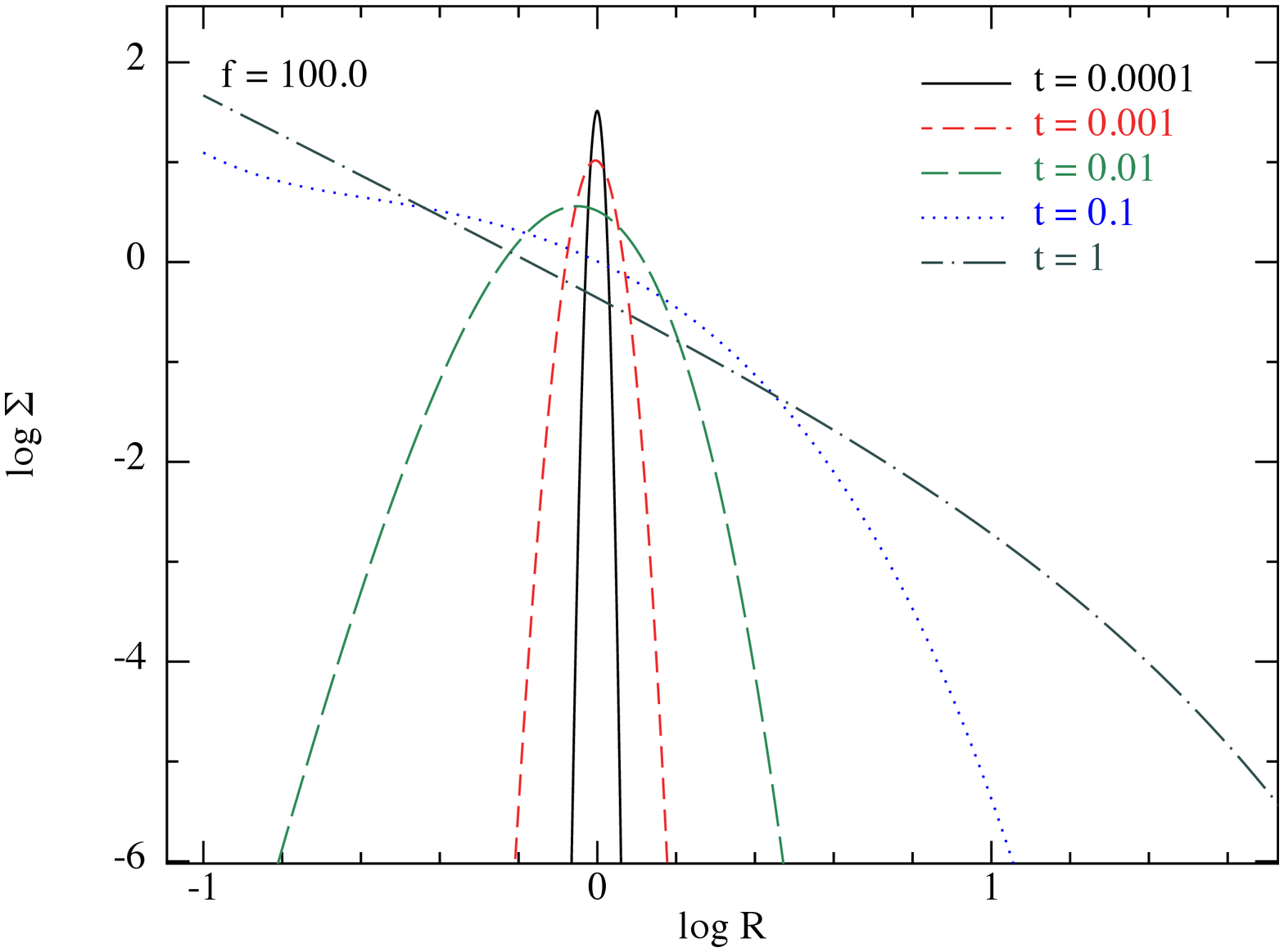}
  \includegraphics[width=0.4\columnwidth]{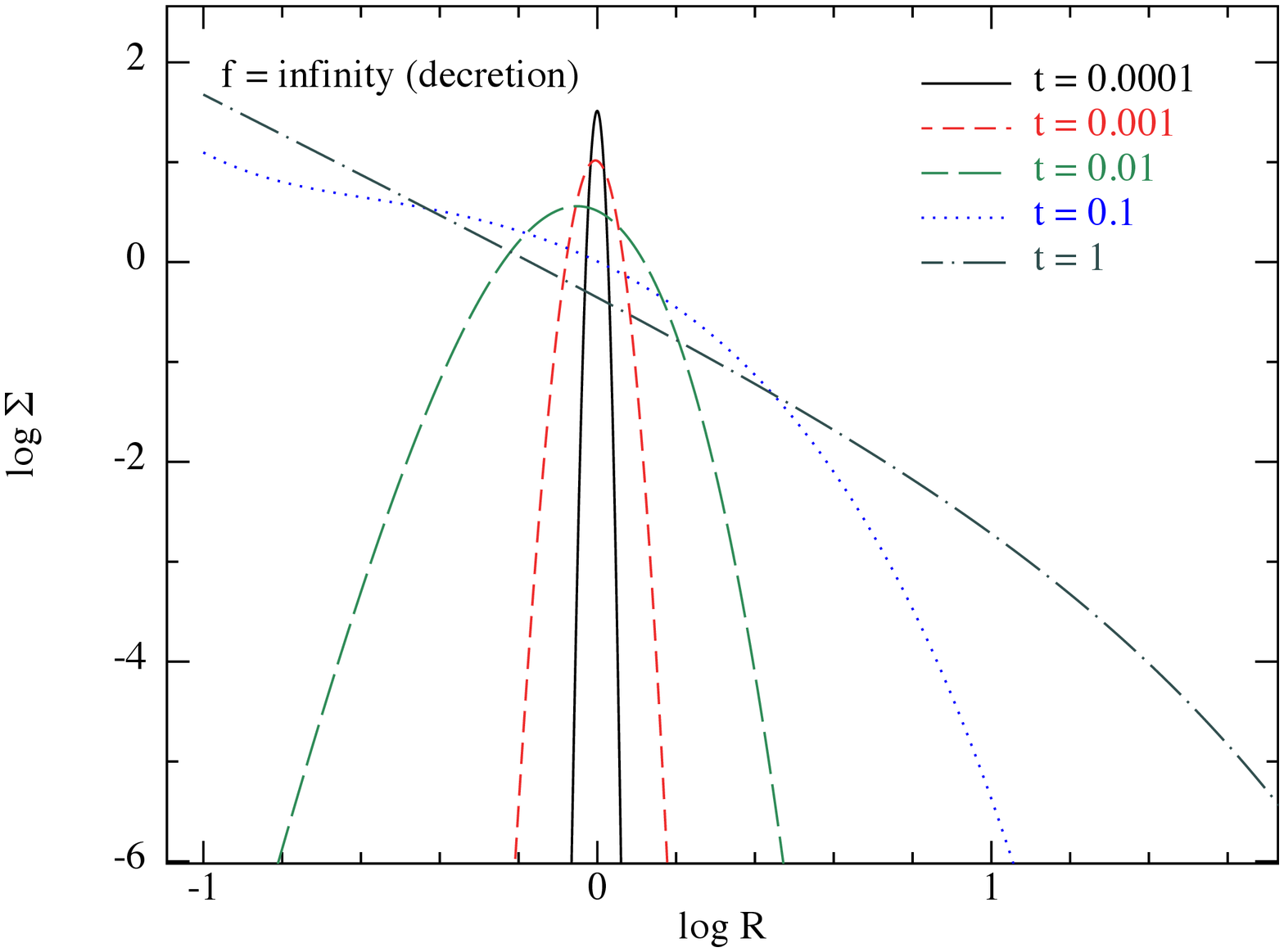}
  \end{center}
  \caption{The same as Fig.~\ref{fig4}, but here with $\nu = kR^{3/2}$.}
  \label{fig6}
\end{figure}

\begin{figure}
  \centering\includegraphics[width=0.5\columnwidth]{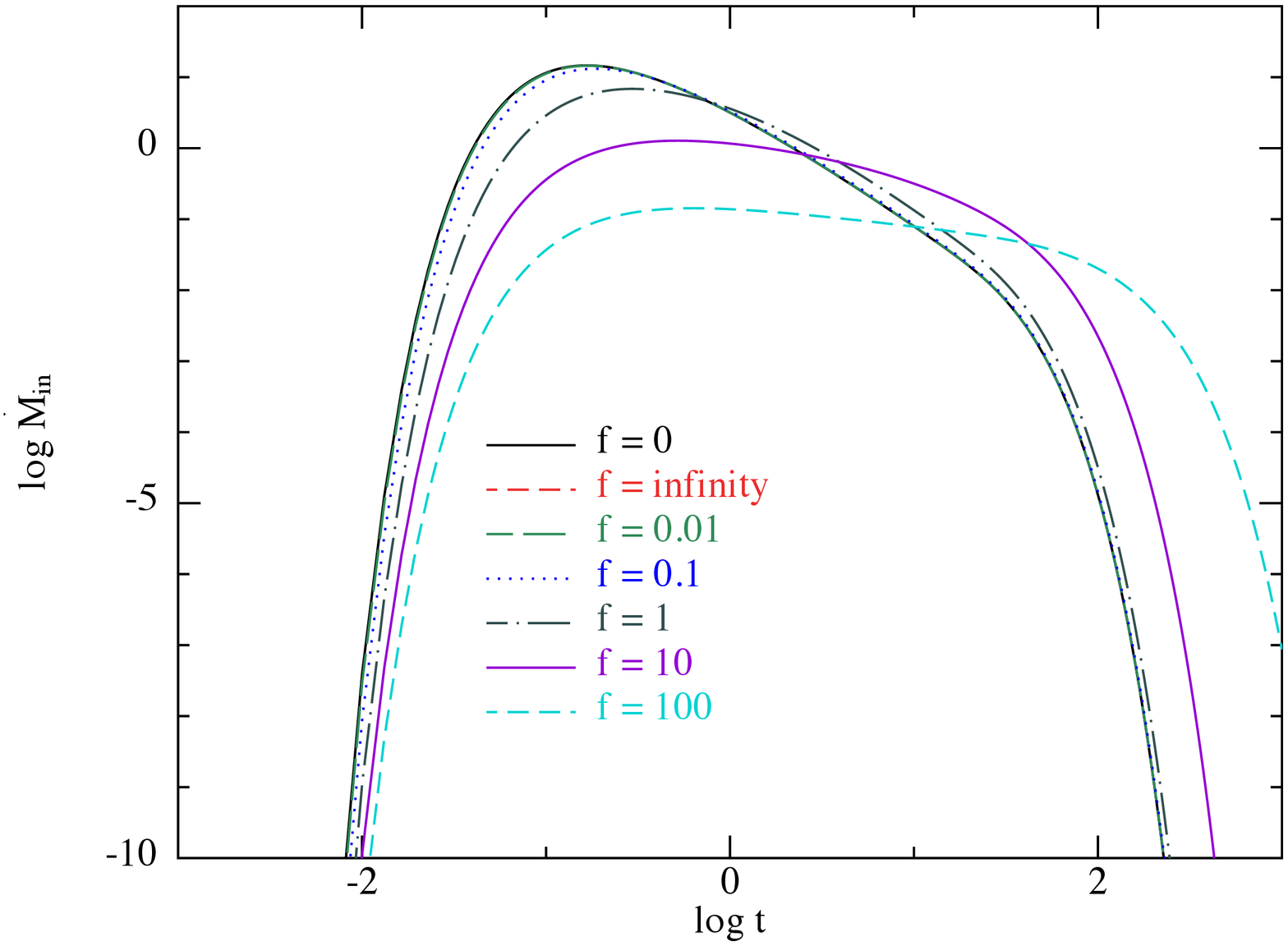}
  \centering\includegraphics[width=0.5\columnwidth]{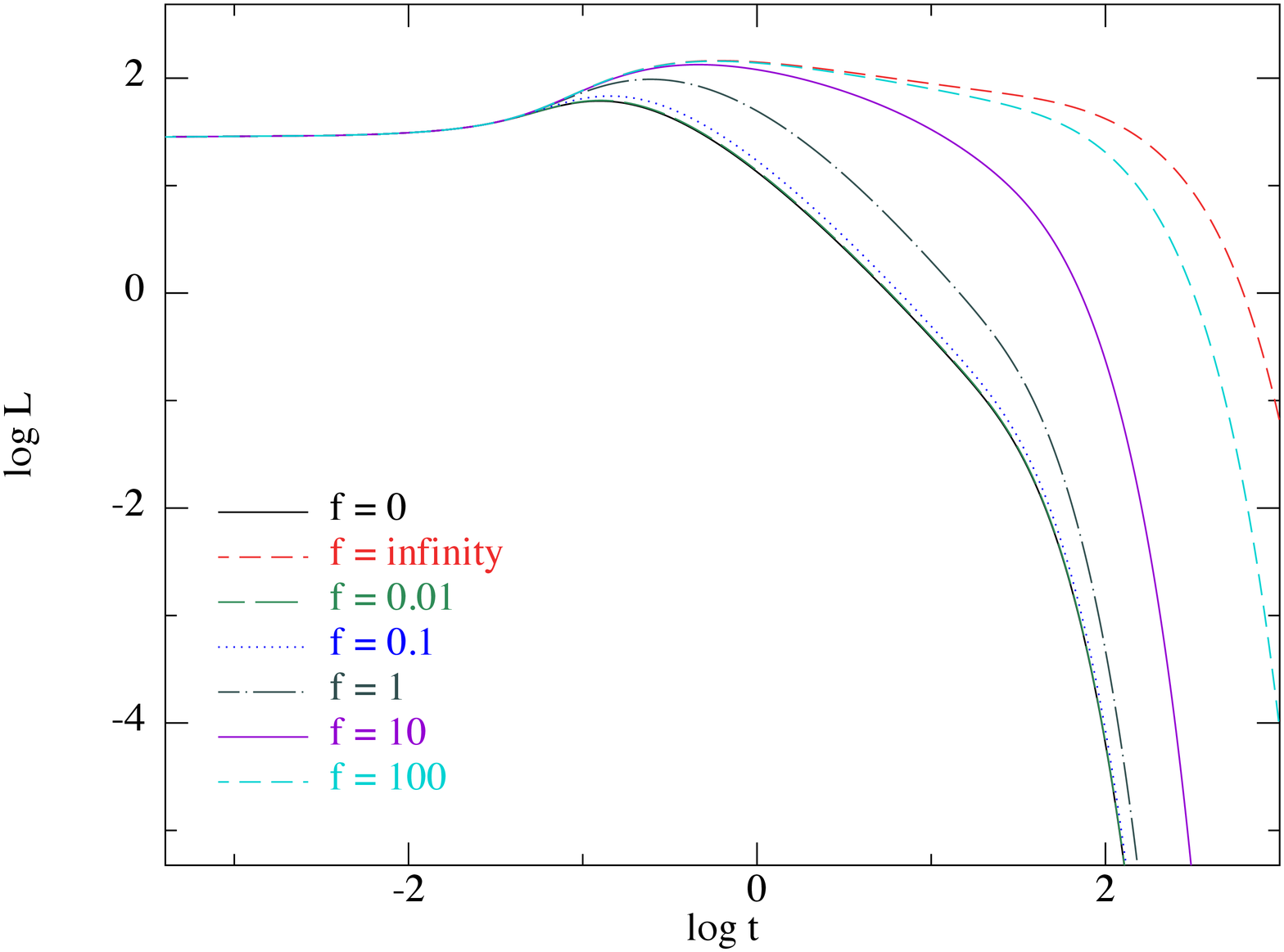}
  \centering\includegraphics[width=0.5\columnwidth]{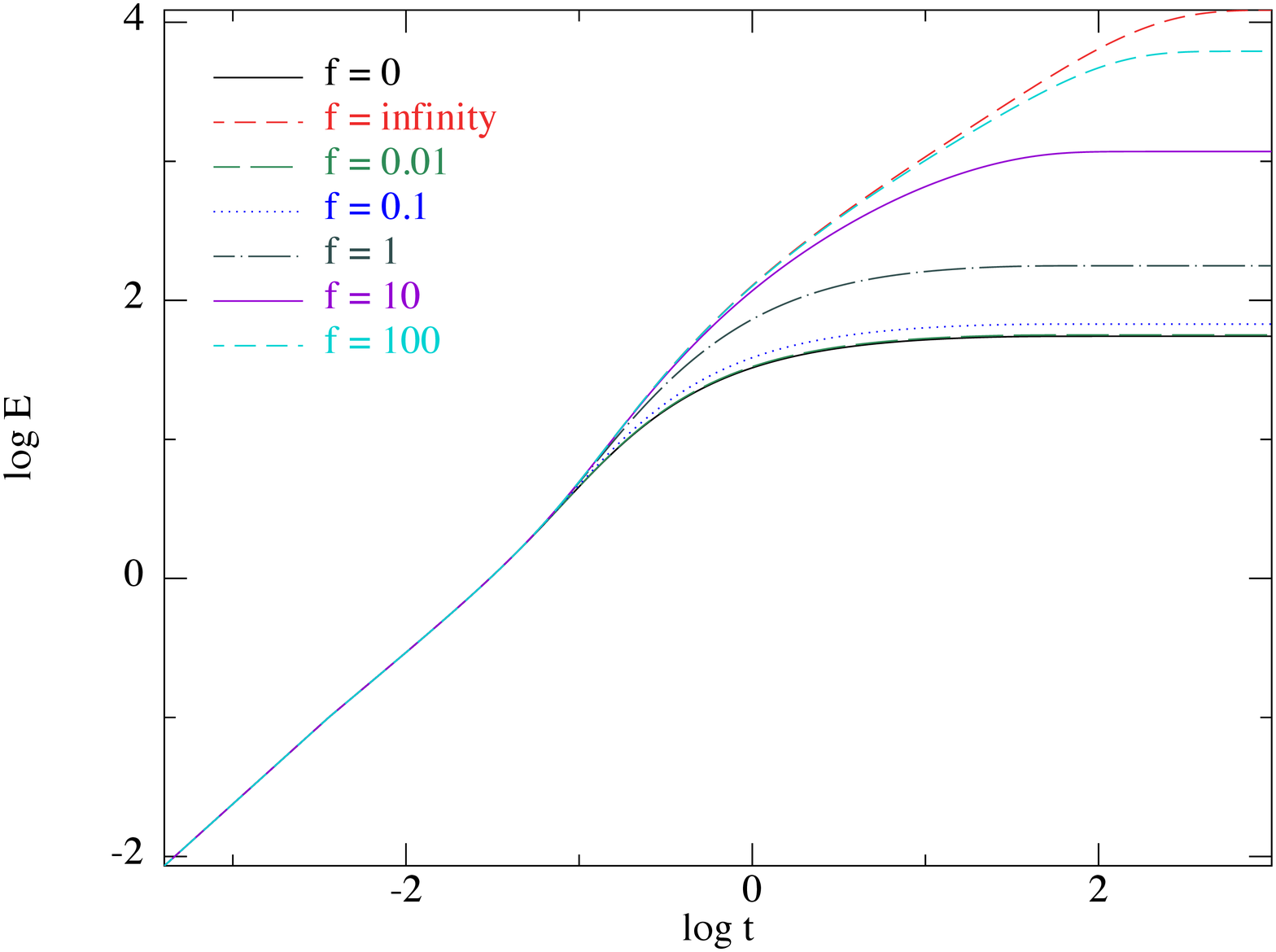}
  \caption{The same as Fig.~\ref{fig5}, but here with $\nu = kR^{3/2}$. In this case the viscous timescale at the outer radius is $t_\nu(R_{\rm out}) = R^{1/2}/k = \sqrt{1000}$. This means that the disc mass drains through the outer boundary much faster than seen in Fig.~\ref{fig5}, and thus the central accretion rate and luminosity fall off sooner in this case.}
  \label{fig7}
\end{figure}

We also provide in Figs.~\ref{fig6} \& \ref{fig7} the results of simulations that are identical to those in Figs~\ref{fig4} \& \ref{fig5} except for the viscosity, which here is $\nu = kR^{3/2}$ rather than $\nu = kR$. In this case, we have $m=0$ and $n=3/2$, which yields ${\dot M}\propto L\propto t^{-2}$ for the accretion disc case and $L$ constant for the decretion disc. This behaviour is evident in these figures, although for the numerical solutions the power-laws are less steep due to matter being lost through the outer boundary. When the value of $f$ is finite, the late time solutions follow the accretion disc case, and thus the time dependence is given by (\ref{latetime}).

These figures demonstrate that accurate numerical simulation of these discs for different viscosity profiles are readily available.

\section{Discussion}
\label{systems}
We have noted that the standard zero inner torque boundary condition ($f=0$) is an adequate approximation for discs around non-magnetic stars for which the stellar angular velocity, $\Omega_\star$, is not close to the break-up speed, $\Omega_K = (GM/R_\star^3)^{1/2}$ \citep{Pringle:1977aa,Pringle:1981aa,Tylenda:1981aa}. If $\Omega_\star$ becomes very close to $\Omega_K$ (assuming that the star remains stable, which it probably does not) then accretion can still continue, but as $\Omega_\star \rightarrow \Omega_K$, the star provides a net torque and $f \rightarrow 1$ \citep{Pringle:1989aa,Popham:1991aa}. In this case (see Equation~\ref{energy}) the disc now radiates three times the gravitational energy that it loses. This extra energy is provided by the stellar rotation, and thus the star is slowed down. This is a particular case in which the inner boundary condition of the disc has a non-zero torque, that is $f \neq 0$. We consider briefly some other cases of astrophysical interest.

\subsection{Discs around black holes}
The early models of accretion discs around black holes used the assumption of zero-torque evaluated at the radius corresponding to the innermost stable circular orbit \citep[ISCO;][]{Shakura:1973aa,Novikov:1973aa,Page:1974aa}. These are the relativistic analogues of the $f=0$ discs.

\cite{Krolik:1999aa}, however, suggested that this boundary condition might need to be amended. He noted that in such discs the gas is assumed to spiral down to the ISCO, and at that point to fall freely inwards. Thus the usual boundary condition assumed there was zero torque at the ISCO. He pointed out that since the viscosity was most likely magnetic in origin, the material falling inwards from the ISCO could in principle transmit angular momentum outwards through magnetic torques. Thus there can be an outward flux of angular momentum at the ISCO, equivalent in a Newtonian sense to having $f > 0$. \cite{Gammie:1999aa} confirms this possibility.

\cite{Paczynski:2000aa} argued that \cite{Krolik:1999aa} was wrong and that there is no effect. In fact \cite{Krolik:1999aa} was right (as Paczy\'nski later confirmed, see \citealt{Afshordi:2003aa}) but the arguments of \cite{Paczynski:2000aa} do show that for a thin disc the effect must be small. The point is that a thin disc has small scale ($\sim H$), where $H \ll R$ is the disc scale-height, dynamically weak ($\varv_A^2 \sim (H/R)^2\varv_\phi^2$) magnetic fields \citep{Shakura:1973aa,Martin:2019aa}. Here $\varv_A$ is the Alfv\`en speed in the disc, and $\varv_\phi = R \Omega$ the disc circular velocity. MHD computations carried out in the pseudo-Newtonian Paczynski-Wiita potential \citep{Armitage:2001aa,Reynolds:2001aa,Hawley:2002aa} find around a 10 per cent effect (i.e. $f \sim 0.1$). \cite{Noble:2010aa} undertake MHD numerical simulations of ``thin'' $(H/R = 0.06 - 0.17)$ discs around a Schwarzschild black hole. They find that net angular momentum accreted per unit rest mass is 7 -- 15  per cent less than the angular momentum at the ISCO. Thus, in Newtonian terms they also find that $f \sim 0.1$.

Recently, the question of finite ISCO stress in thin relativistic accretion discs has been revisited by \cite{Balbus:2018aa} and \cite{Mummery:2019aa,Mummery:2019ab,Mummery:2020aa,Mummery:2020ab}. Their primary aim is to try to obtain fits to the long-term light curves of some tidal disruption events (TDEs) which sometimes show that the luminosity as a function of time varies as $L(t) \propto t^{-n}$ with $n \approx 0.7$ \citep{Auchettl:2017aa}. The disc evolution solutions they present are essentially Green's functions, since they assume that initially a tidally disrupted star is deposited in circular orbits at a radius of a few times the ISCO (as opposed to the typically assumed $t^{-5/3}$ fallback for full disruptions and $t^{-9/4}$ fallback for partial disruptions; \citealt{Coughlin:2019aa,Miles:2020aa}). They then obtain fits to the light curves by adjusting the inner boundary conditions at the ISCO (in particular they vary a parameter $\gamma$, where $1/\gamma$ is basically the same as $f$ with some differences resulting from using Newtonian gravity as opposed to general relativity; Balbus, private communication).

\cite{Mummery:2019ab} consider objects which at intermediate times show shallow power-law luminosity behaviour with $L\propto t^{-n}$ where $n \sim 0.7$. To obtain their solutions, they have chosen Green's functions corresponding to values of $f \gg 1$ (typically, in Newtonian terms, we estimate through for example comparison of Fig~\ref{fig3} here with Fig.~1 of \citealt{Mummery:2019ab} that they require $f \sim 10-100$). As we have seen, such large values of $f$ imply that most of the energy radiated by the disc is provided by the central gravitating object. However the models of \cite{Mummery:2019ab} consider accretion onto a central (non-rotating) black hole. How such a black hole is able to provide prolonged power, which in total exceeds the available gravitational energy of the initial ring of material by one to two orders of magnitude, is not addressed.

\subsection{Discs around rotating magnetic stars}
If the central object in an accretion disc is a rotating magnetic star, then the inner edge of the disc may be defined by the strength of the magnetic field. \cite{Pringle:1972aa}, in their model for pulsing binary X-ray sources,  identified two radii relevant to this process. First, there is the co-rotation radius, $R_\Omega$ at which the Keplerian angular velocity is equal to that of the star, $R_\Omega = (GM/\Omega_\star^2)^{1/3}$, and second the magnetospheric radius, $R_M$ at which the stellar magnetic field is strong enough to disrupt the disc flow. They argued that if the star is rotating slowly, in the sense that $R_\Omega > R_M$ then, once disrupted, the disc material can attach itself to the field lines and accrete onto the central star. In this case the star is spun up and the stellar rotation period changes at a rate determined by the accretion rate and the lever arm $R_M$. However, if the star is a fast rotator, in the sense that $R_M > R_\Omega$ then although the disc is disrupted, the disc material is flung outwards by centrifugal force and does not accrete. In this case the star is spun down (see for example \citealt{Benli:2020aa}). As an example, this behaviour is thought to provide energy injection into surrounding matter following a supernova that yields a magnetar \cite[][]{Piro:2011aa}.

The exact value of $R_M$ in any particular case is not straightforward to determine, as can be seem from the various different suggestions as to how it may be determined \citep[e.g.][]{Pringle:1972aa,Bath:1974aa,Ghosh:1978aa,Ghosh:1979aa,Ghosh:1979ab}, and by the complexity of the disc-magnetosphere interactions seen in numerical simulations \citep{Miller:1997aa,Romanova:2003aa,Romanova:2003ab}. However, since the estimates involve the magnetic energy density which, for a stellar dipole field, falls off with a high power of radius, $B^2 \propto R^{-6}$, the various computed values of $R_M$ do not differ greatly.

For fast rotators, with $R_M \gg R_\Omega$ we may expect that accretion is essentially prevented and $f \gg 1$. This is presumably the case for the discs around Be stars, normally referred to as ``decretion discs''. The disc formation mechanism, and thus the inner boundary for these discs remains unknown \citep{Rivinius:2013aa}. However, \cite{Nixon:2020aa} have argued that the formation of the disc and the prevention of re-accretion of disc material is caused by small scale magnetic fields in the atmosphere of the fast rotating Be star.

The spin history of young stellar objects (T Tauri stars) can be modelled in terms of magnetospheric disc accretion \citep[e.g.][]{Matt:2010aa}, although for these objects the field is dynamo produced and therefore likely to be variable \citep{Clarke:1995aa}. For these objects the spin up/spin down rate is typically less than the disc lifetimes, and so one expects to find typically that $R_M \approx R_\Omega$. In such a case, where $d\Omega_\star/dt \approx 0$, the specific angular momentum of material being accreted by the star ($R_\star^2 \Omega_\star$) is less than the specific angular momentum of the matter arriving at the inner disc edge ($\approx R_\Omega^2 \Omega_\star$). This implies that most of the angular momentum arriving at the inner disc edge is not absorbed by the star. To conserve angular momentum this implies that there must be an outward flux of angular momentum at the inner disc edge which must be such that $f \approx 1 - (R_\star/R_\Omega)^2$.

More generally, if $R_M < R_\Omega$, so the accretion occurs without problem (as in the case of the pulsing binary X-ray sources discussed by \citealt{Pringle:1972aa}) similar arguments imply that in this case too $f > 0$. The exact value of $f$ to be applied here depends on the details of the flow in the region where the disc and the stellar magnetosphere interact, but again, to conserve angular momentum it must be roughly such that $f \approx 1 - (R_M/R_\Omega)^2$. 

\subsection{Discs around binary systems} 
For circumbinary discs the central binary, whether consisting of two stars \citep[e.g.][]{Lin:1979aa,Artymowicz:1991aa} or two black holes \citep[e.g.][]{Begelman:1980aa}, provides a source of angular momentum and energy. Resonances between the disc and binary orbits (in this case outer Lindblad resonances) transfer energy and angular momentum from the binary to the disc \citep{Lin:1986aa,Pringle:1991aa,Artymowicz:1994aa}. In general this leads to the launching of a wave through the disc \citep{Lubow:1998aa}, and depending on the disc parameters the wave may propagate outwards or locally deposit its energy and angular momentum \citep[see the discussion in][]{Heath:2020aa}.

If significant amounts of energy and angular momentum can be transferred to the disc orbits at small radii (of order several times the binary semi-major axis) then the disc can be efficiently truncated. In this case, which typically occurs when the disc viscosity is not large, the accretion rate on to the binary can be significantly reduced with only a small amount of matter leaking on to the binary through time-dependent streams \citep[e.g.][]{Artymowicz:1991aa,Artymowicz:1994aa,Heath:2020aa}. Here we expect $f \gg 1$. For discs in which the viscous torque is comparable to the torque applied to the disc by resonances, a non-axisymmetric cavity can form around the binary, with the binary fed mass by time-dependent streams \citep{Artymowicz:1996aa}, in which case we expect $f \sim 1$.  For sufficiently high viscosity (which typically requires the disc to be thick), there may be no resonance in the disc that can arrest the accretion flow and matter can accrete unimpeded on to the binary. In this case there is still energy and angular momentum transferred from the binary to the disc by resonances, but the effect is substantially reduced as there is no significant build up of matter at the resonance locations. Therefore, in this case we expect $f \ll 1$. The exact value of $f$ in each case is dependent on several parameters which interplay in a complex manner \citep[see][for a broad discussion on the topic]{Heath:2020aa}.

The binary system has a limited supply of energy and angular momentum that can be fed into the disc. Once the torque applied at the inner edge has been applied for long enough, that the transferred energy and angular momentum becomes comparable with the initial binary orbit, the binary orbit will have evolved significantly. In this case, the binary semi-major axis typically shrinks with time due to the loss of energy, and the eccentricity may grow or decay with time depending on the inner disc structure with more asymmetric structures typically leading to greater eccentricity growth.

\section{Conclusions}
\label{conclusions}
We have presented analytical and numerical calculations of accretion discs with a non-zero torque inner boundary condition. We have defined the parameter $f$ as the ratio of the outward viscous flux of angular momentum to the inward advected flux at the inner disc boundary. Our results approach the standard cases of accretion discs ($f=0$) and decretion discs ($f\rightarrow\infty$) in the appropriate limits. For $f > 0$ both energy and angular momentum are fed into the disc through the inner disc boundary, and we have noted that for large values of $f$ the energy emitted by the disc is dominated by the energy provided by the central object. We have provided a numerical scheme that accurately reproduces the analytical solutions, and is amenable to more general accretion disc simulations than those presented here. For example, this can be used to simulate discs where the viscosity has an explicit time dependence and/or where the viscosity depends on other disc variables such as the disc surface density.

We have discussed several astrophysical systems where a non-zero central torque is present. For discs around black holes, it is generally accepted that the central torque---driven by magnetic torques acting between disc matter that is connected across the ISCO---is small and has little impact on the disc structure. Comparison between our disc solutions and the values of $f$ inferred from MHD simulations in the literature ($f \lesssim 0.1$) confirms this view. We have noted that this is in direct conflict with the models of TDE light-curves by \cite{Mummery:2019ab} which make the assumption that $f \sim 10 - 100$. For discs around magnetic stars or binary systems, our methodology provides a framework in which investigations (analytical or numerical) may more accurately represent the physics of the disc inner boundary condition. This may be achieved by applying the boundary condition (\ref{innerBC}) with an appropriate value of $f$, which may itself depend on the local physical setup and disc properties, instead of switching instantly between accretion ($f=0$) and decretion ($f=\infty$). For example, the long term evolution of supermassive black hole binaries interacting with circumbinary discs may be modelled by taking account of the changing disc conditions---and thus the $f$ value---as the binary orbit evolves, and \cite{Nixon:2020aa} suggest that the lightcurves of Be stars may be modelled by a time-dependent $f$ that varies with the evolution of small scale magnetic fields at the stellar surface. 

\section*{Acknowledgements}
We thank the referee, Stephen Lubow, for helpful input. We thank Steven Balbus for useful correspondence. We thank Eric Coughlin for useful comments on the manuscript. CJN is supported by the Science and Technology Facilities Council (grant number ST/M005917/1). CJN acknowledges funding from the European Union’s Horizon 2020 research and innovation program under the Marie Sk\l{}odowska-Curie grant agreement No 823823 (Dustbusters RISE project). This research used the ALICE High Performance Computing Facility at the University of Leicester. This work was performed using the DiRAC Data Intensive service at Leicester, operated by the University of Leicester IT Services, which forms part of the STFC DiRAC HPC Facility (\url{www.dirac.ac.uk}). The equipment was funded by BEIS capital funding via STFC capital grants ST/K000373/1 and ST/R002363/1 and STFC DiRAC Operations grant ST/R001014/1. DiRAC is part of the National e-Infrastructure.

\bibliographystyle{aasjournal}
\bibliography{nixon}

%%%%%%%%%%%%%%%%% APPENDICES %%%%%%%%%%%%%%%%%%%%%
\appendix
\section{Time-dependent discs with general inner boundary conditions}
\label{appA}
We present in Section~\ref{timedep} the solution for $\sigma(x,t)$ of the time dependent disc with inner radius $R_{\rm in}$, outer radius at infinity, viscosity $\nu=kR$, and initial profile $\sigma(x,t=0) = \sigma_0\delta(x-x_{\rm add})$. In this appendix we provide the derivation and details. As noted above the solution (\ref{Greengeneral}) is derived for the specific case of $\nu = kR$ with $k$ a constant. For more general cases, we show in Section~\ref{numerical} that accurate solutions can be obtained readily following a simple numerical scheme \cite[e.g.][]{Pringle:1991aa}.

\subsection{General solution for finite disc}
First we consider the case of a finite disc from $R_{\rm in}$ to $R_{\rm out}$. We have $\nu = kR$ and therefore
\begin{equation}
\sigma = k^{-1} (\nu \Sigma R^{1/2}) = \Sigma R^{3/2} = S/k\,.
\end{equation}

We also define $x = R^{1/2}$ and the evolution equation for surface density then becomes
\begin{equation}
\label{sigmaequation}
\frac{\partial\sigma}{\partial t} = c^2 \frac{\partial^2\sigma}{\partial x^2}\,,
\end{equation}
where $c^2 = 3k/4$.

At the outer boundary $x = x_{\rm out}$ we take $\sigma = 0$. At the inner boundary the condition (\ref{innerBC}) becomes
\begin{equation}
\label{innerBC2}
\sigma-fx\frac{\partial\sigma}{\partial x} = 0\,,
\end{equation}
evaluated at $x = x_{\rm in}$. 

To solve this, we first separate variables and write $\sigma(r,t) = X(x)\, T(t)$. Then we find $T \propto \exp (-\lambda^2 t)$, and that in general
\begin{equation}
\label{generalX}
X(x) = A\cos\left(\lambda(x-x_{\rm in})/c\right)+B\sin\left(\lambda(x-x_{\rm in})/c\right)\,.
\end{equation}

Then applying the boundary conditions at $x=x_{\rm in}$, $x_{\rm out}$ we find that
\begin{equation}
X(x) = A_n\phi_n(x)\,,
\end{equation}
where the eigenfunctions, for $n = 1,2,3, \ldots$, are 
\begin{equation}
\phi_n(x) = \sin\left[\lambda_n(x-x_{\rm in})/c\right]+\lambda_n\left(fx_{\rm in}/c\right)\cos\left[\lambda_n(x-x_{\rm in})/c\right]\,,
\end{equation}
and the eigenvalues $\lambda_n > 0, \: n = 1,2,3, \ldots$ are defined as the positive roots of
\begin{equation}
\tan\left[\lambda\left(x_{\rm out}-x_{\rm in}\right)/c\right] = -(fx_{\rm in}/c)\lambda\,,
\end{equation}
with $\lambda_1 < \lambda_2 < \lambda_3 \ldots$.

We also note the orthogonality relation
\begin{equation}
\label{norm}
\int_{x_{\rm in}}^{x_{\rm out}}\phi_m\phi_n\,dx = C_n\delta_{nm}\,,
\end{equation}
with
\begin{equation}
C_n = \frac{1}{2}(x_{\rm out}-x_{\rm in})\left[1+\lambda_n^2\left(\frac{fx_{\rm in}}{c}\right)^2\right]+\frac{1}{2}fx_{\rm in}\,.
\end{equation}

Thus the general solution for $\sigma$ is given by
\begin{equation}
\label{gensolution}
\sigma(x,t) = \sum_{n=1}^{\infty} A_n\phi_n(x)\exp(-\lambda_n^2 t)\,,
\end{equation}
where the $A_n$ are determined from the initial conditions.

\subsubsection{Green's function}
For example, for the Green's function we set
\begin{equation}
\label{sigmazero}
\sigma(x,t=0) = \sigma_0\,\delta(x-x_{\rm add}),
\end{equation}
with $x_{\rm in} < x_{\rm add} < x_{\rm out}$.

Substituting this into the general solution and using the orthogonality relations we find that the Green's function is given by (\ref{gensolution}) with the coefficients defined by
\begin{equation}
A_n = \sigma_0 \phi_n(x_{\rm add})/C_n,
\end{equation}
where the $C_n$ are defined by (\ref{norm}).

We note that at late times the first term of the sum dominates and thus the solution is of the form $\sigma \propto \exp(- \lambda_1^2 t)$. For the case $x_{\rm out} \gg x_{\rm in}$ it is straightforward to show that $\lambda_1^2 = (4/3 Y_1)^2 (\nu(R_{\rm out})/R_{\rm out}^2)$ where $Y_1$ is the smallest positive solution of $ \tan Y = f (x_{\rm in}/x_{\rm out}) Y$. We can see that $\pi/2 < Y_1 < \pi$ with $Y_1 \approx \pi/2$ when $f \ll x_{\rm in}/x_{\rm out}$ and $Y_1 \approx \pi$ when $f \gg x_{\rm in}/x_{\rm out}$.

\subsection{General solution for the infinite disc}
Now, we consider the case of an infinite disc, where the inner edge is at $R_{\rm }$ and the outer edge is at $R_{\rm out} = \infty$. Again we shall look for the Green's function, which is the solution to (\ref{sigmaequation}) with initial $\sigma-$distribution, $\sigma(x,t=0) = \sigma_0\delta(x-x_{\rm add})$.

As above we separate variables in the form $\sigma(x,t)\propto\exp(-\lambda^2 t)X(x)$. Applying the boundary condition at $x = x_{\rm in}$ we can then write the general solution as
\begin{eqnarray}
\label{sigmax}
\sigma(x,t) & = & \int_0^\infty \exp(-\lambda^2 t)\,B_\lambda\left\{\sin\left[\lambda (x- x_{\rm in})/c\right]\right. \\
& & + \left.(fx_{\rm in}/c)\lambda\cos\left[\lambda(x-x_{\rm in})/c\right]\right\}\,{\rm d}\lambda\,, \nonumber
\end{eqnarray}
where $B_\lambda$ is to be determined from the initial conditions.

We now wish to proceed by means of Fourier sine and cosine transforms. To simplify the analysis it is convenient to substitute
\begin{equation}
y = (x-x_{\rm in})/c \ge 0\,,
\end{equation}
so that $x-x_{\rm in}+cy$ and we write $x_{\rm add} = x_{\rm in}+cy_{\rm add}$.

Then we can rewrite (\ref{sigmax}) as
\begin{equation}
\label{sigmay}
\sigma(y,t) = \int_0^\infty \exp(-\lambda^2 t) B_\lambda\left\{\sin(\lambda y)+(fx_{\rm in}/c)\lambda\cos(\lambda y)\right\}\,{\rm d}\lambda\,.
\end{equation}

At time $t = 0$ we now let $\sigma(y,t=0) = X_0(y)$. Then we have that
\begin{equation}
X_0(y) = \int_0^\infty B_\lambda [\sin(\lambda y)+(fx_{\rm in}/c)\lambda\cos(\lambda y)]\,{\rm d}\lambda\,,
\end{equation}
and
\begin{equation}
\frac{{\rm d}X_0(y)}{{\rm d}y} = \int_0^\infty B_\lambda[\lambda\cos(\lambda y)-(fx_{\rm in}/c)\lambda^2\sin(\lambda y)]\,{\rm d}\lambda\,.
\end{equation}

Hence we find that
\begin{equation}
X_0-\frac{fx_{\rm in}}{c}\frac{{\rm d}X_0}{{\rm d}y} = \int_0^\infty B_\lambda\left[1+\left(\frac{fx_{\rm in}}{c}\right)^2\lambda^2\right]\sin(\lambda y)\,{\rm d}y\,.
\end{equation}

We can then invert the sine-transform to obtain
\begin{equation}
B_\lambda\left[1+\left(\frac{fx_{\rm in}}{c}\right)^2\lambda^2\right] = \frac{2}{\pi}\int_0^\infty\left[X_0-\left(\frac{fx_{\rm in}}{c}\right)\frac{{\rm d}X_0}{{\rm d}y}\right]\sin(\lambda y)\,{\rm d}y\,.
\end{equation}

Recalling that from the properties of the Fourier sine/cosine transforms 
\begin{equation}
\int_0^\infty\frac{{\rm d}X_0}{{\rm d}y}\sin(\lambda y)\,{\rm d}y = -\lambda\int_0^\infty X_0\cos(\lambda y)\,{\rm d}y\,,
\end{equation}
we can then write $B_\lambda$ in terms of the initial condition $X_0(y)$, viz.
\begin{equation}
B_\lambda \left[1+\left(\frac{fx_{\rm in}}{c}\right)^2\lambda^2\right] = \frac{2}{\pi}\int_0^\infty X_0(y)\sin(\lambda y)\,{\rm d}y+\frac{2}{\pi}\left(\frac{fx_{\rm in}}{c}\right)\lambda\int_0^\infty X_0(y)\cos(\lambda y)\,{\rm d}y\,.
\end{equation}

\subsubsection{Green's function}
For the Green's function we insert the initial condition (\ref{sigmazero}) which here becomes
\begin{equation}
X_0(y) = (\sigma_0/c)\delta(y-y_{\rm add})\,,
\end{equation}
where we note that $y_{\rm add} > 0$.

In this case we obtain
\begin{equation}
B_\lambda = \left\{\frac{2\sigma_0}{\pi c}\sin(\lambda y_{\rm add})+\frac{2\sigma_0}{\pi c}\left(\frac{fx_{\rm in}}{c}\right)\lambda\cos(\lambda y_{\rm add})\right\}\left[1+\left(\frac{fx_{\rm in}}{c}\right)^2\lambda^2\right]^{-1}\,.
\end{equation}

We then substitute this expression for $B_\lambda$ into (\ref{sigmay}) to obtain $\sigma(y,t)$. 
\medskip
In order to evaluate the integral we need the following three identities from \cite{Gradshteyn:1980aa}:
\begin{enumerate}
\item (p. 497)
\begin{eqnarray}
\int_0^\infty \exp(-\beta x^2)\sin(ax)\frac{x\,{\rm d}x}{\gamma^2+x^2} & = & -\frac{\pi}{4}\exp(\beta\gamma^2) \\
& & \times\left[2\sinh(a\gamma)+e^{-a\gamma}{\rm erf}\left(\gamma\sqrt{\beta}-\frac{a}{2\sqrt{\beta}}\right)\right. \nonumber \\
& & -\left.e^{a\gamma}{\rm erf}\left(\gamma\sqrt{\beta}+\frac{a}{2\sqrt{\beta}}\right)\right]\,, \nonumber
\end{eqnarray}
valid for $\Re(\beta) > 0$, $\Re(\gamma) > 0$ and $a > 0$.

\item (p. 497)
\begin{eqnarray}
\int_0^\infty \exp(-\beta x^2)\cos(ax)\frac{{\rm d}x}{\gamma^2+x^2} & = & -\frac{\pi}{4\gamma}\exp(\beta \gamma^2) \\
& & \times\left[2\cosh(a\gamma)-e^{-a\gamma}{\rm erf}\left(\gamma\sqrt{\beta}-\frac{a}{2\sqrt{\beta}}\right)\right. \nonumber \\
  & & \left. -e^{a\gamma}{\rm erf}\left(\gamma\sqrt{\beta}+\frac{a}{2\sqrt{\beta}}\right)\right]\,, \nonumber
\end{eqnarray}
valid similarly, and

\item (p. 480)
\begin{equation}
\int_0^\infty \exp(-\beta x^2)\cos(bx)\,{\rm d}x = \frac{1}{2}\sqrt{\frac{\pi}{\beta}}\exp\left(\frac{-b^2}{4\beta}\right),
\end{equation}
valid for $\Re(\beta) > 0$.

\end{enumerate}

\medskip

Here ${\rm erf}(x)$ is the error function which is defined as
\begin{equation}
\label{error}
{\rm erf}(x) = \frac{2}{\sqrt{\pi}}\int_0^x e^{-t^2}\,{\rm d}t\,,
\end{equation}
and the complimentary error function is
\begin{equation}
\label{errorc}
{\rm erfc}(x) = 1-{\rm erf}(x)
\end{equation}
Making use of these we obtain the Green's function which we give in (\ref{Greengeneral}).

\section{Evaluating the Green's function}
\label{estGreen}
The Green's function for a general disc (\ref{Greengeneral}) is
\begin{eqnarray}
\label{Greengeneral2}
\sigma(x,t) & = & \frac{\sigma_0}{2c\sqrt{\pi t}}\times\left\{\exp\left[-\frac{(x-x_{\rm add})^2}{4c^2t}\right]+\exp\left[-\frac{(x+x_{\rm add}-2x_{\rm in})^2}{4c^2t}\right]\right\} \nonumber \\
&& -\,\frac{\sigma_0}{fx_{\rm in}}\exp\left(\frac{c^2t}{f^2x_{\rm in}^2}\right)\exp\left[\frac{x+x_{\rm add}-2x_{\rm in}}{fx_{\rm in}}\right]{\rm erfc}\left(\frac{c\sqrt{t}}{fx_{\rm in}}+\frac{x+x_{\rm add}-2x_{\rm in}}{2c\sqrt{t}}\right)\,.
\end{eqnarray}  
Once the argument of ${\rm erfc}$ becomes large, and thus the argument of the first exponential function in the second line becomes large, this equation can become difficult to evaluate numerically. Here we note that when the argument of ${\rm erfc}$ becomes large, we can replace it with its asymptotic form, given by
\begin{equation}
  {\rm erfc}(A) = \frac{\exp(-A^2)}{A\sqrt{\pi}}\chi(A)
\end{equation}
where $\chi(A)$ is given by the asymptotic series
\begin{equation}
  \chi(A) \sim 1 + \sum_{i=1}^{\infty}(-1)^{i} \frac{1.3.5\ldots(2i-1)}{(2A^2)^i}\,.
\end{equation}

Using this, we can write
\begin{eqnarray}
\label{estGreengeneral}
\sigma(x,t) & = & \frac{\sigma_0}{2c\sqrt{\pi t}}\times\left\{\exp\left[-\frac{(x-x_{\rm add})^2}{4c^2t}\right]+\exp\left[-\frac{(x+x_{\rm add}-2x_{\rm in})^2}{4c^2t}\right]\right\} \nonumber \\
&& -\,\frac{\sigma_0}{fx_{\rm in}}\frac{1}{A\sqrt{\pi}}\exp\left(-\frac{x+x_{\rm add}-2x_{\rm in}}{4c^2t}\right)\chi(A)\,,
\end{eqnarray}  
where $A$ is the argument of ${\rm erfc}$ in (\ref{Greengeneral2}), i.e.,
\begin{equation}
  A = \frac{c\sqrt{t}}{fx_{\rm in}}+\frac{x+x_{\rm add}-2x_{\rm in}}{2c\sqrt{t}}\,.
\end{equation}

We have found that, for example, when $A \gtrsim 100$, then a quadruple precision floating point arithmetic overflow is caused for the term $\exp(c^2t/f^2x_{\rm in}^2)$. At this point, one may replace (\ref{Greengeneral2}) with (\ref{estGreengeneral}) and evaluate $\chi(A) \approx \chi(A,n)$ where
\begin{equation}
  \chi(A,n) = 1 + \sum_{i=1}^{n}(-1)^{i} \frac{1.3.5\ldots(2i-1)}{(2A^2)^i}\,.
\end{equation}
Then an iteration can be performed over $n$ to determine a converged answer to the required level of precision. We find that for $A \gtrsim 100$, then $n \lesssim 10$ is sufficient to maintain accuracy to quadruple precision.

%%%%%%%%%%%%%%%%%%%%%%%%%%%%%%%%%%%%%%%%%%%%%%%%%%
\end{document}